\documentclass[letterpaper, abstracton, DIV11, 12pt, titlepage=false]{scrartcl}

\usepackage{amsmath,amsfonts,amssymb,bbm} 
\usepackage[round]{natbib}
\usepackage{microtype}
\usepackage{pifont}
\usepackage{dsfont}
\usepackage{setspace}
\onehalfspace
\usepackage[usenames,dvipsnames,svgnames,table]{xcolor}
\usepackage{tikz}
\usetikzlibrary{matrix,positioning}
\usepackage{amsthm}
\usepackage{float}
\usepackage{setspace}
\usepackage{numprint}
\npstyleenglish
\npthousandsep{\rq}
\usepackage{pgfplots}
\usepgfplotslibrary{groupplots}
\pgfplotsset{compat=newest}
\usepackage{bm}
\usepackage{mathabx}
\usepackage{mathabx}
\usepackage[colorlinks=true,bookmarks=true, pdftex, citecolor = blue, linkcolor = blue,urlcolor = blue]{hyperref}

\DisableLigatures[f]{encoding = *, family = * }

\def\bf{\normalfont\bfseries}
\changenotsign

\newcommand{\cmark}{{\color{OliveGreen}\ding{51}}}
\newcommand{\xmark}{{\color{BrickRed}\ding{55}}}

\newcommand{\DA}{\text{DA}}

\newcommand{\ABM}{\text{ABM}}
\newcommand{\BM}{\text{BM}}
\newcommand{\RSD}{\text{RSD}}

\newcommand{\PP}{{\mathds{P}}}
\newcommand{\UU}{{\mathds{U}}}
\newcommand{\pp}{{\pi}}

\newcommand{\DAU}{\DA^\UU}
\newcommand{\ABMU}{\ABM^\UU}
\newcommand{\BMU}{\BM^\UU}
\newcommand{\DAP}{\DA^\PP}
\newcommand{\ABMP}{\ABM^\PP}
\newcommand{\BMP}{\BM^\PP}
\newcommand{\DApi}{\DA^\pp}
\newcommand{\ABMpi}{\ABM^\pp}
\newcommand{\BMpi}{\BM^\pp}

\newtheorem{claim}{Claim}

\theoremstyle{plain}
\newtheorem{theorem}{Theorem}
\newtheorem{lemma}{Lemma}
\newtheorem{proposition}{Proposition}

\theoremstyle{definition}
\newtheorem{definition}{Definition}

\newtheorem{example}{Example}
\newcommand{\ourrep}{}

\theoremstyle{remark}
\newtheorem{remark}{Remark}

\newcommand{\ThmNbmDomDAStatement}{%
For any priority distribution $\mathds{P}$ and any preference profile $ P\in \mathcal{P}^N$, 
if $\BM^{\mathds{P}}( P)$ and $\DAP( P)$ are comparable by rank dominance at $ P$, 
then $\BM^{\mathds{P}}( P)$ rank dominates $\DAP( P)$ at $ P$. 
}
\newcommand{\ThmNbmDomDAProof}{%
We first establish the following lemmas about rank dominance. 
Let $x,x^1,\ldots,x^K \in \Delta(X)$ be assignments such that 
\begin{equation}
	x = \sum_{k = 1}^K x^k \cdot \alpha_k 
\end{equation}
for some $\alpha_1, \ldots, \alpha_K > 0$ with $\sum_{k=1}^K \alpha_k = 1$, i.e., $x$ is the convex combination of the assignments $x^k,k=1,\ldots,K$ with coefficients $\alpha_k,k=1,\ldots,K$. 
\begin{lemma}
\label{lem:rank_distribution_of_convex_combinations} 
The rank distribution $d^x$ of $x$ (at some preference profile $ P$) is equal to the convex combination of the rank distributions $d^{x^k}$ of the $x^k$ with respect to coefficients $\alpha_1,\ldots,\alpha_K$, i.e., 
\begin{equation}
	d^x = \sum_{k=1}^K d^{x^k} \cdot \alpha_k.
\end{equation}
\end{lemma}
Lemma \ref{lem:rank_distribution_of_convex_combinations} is obvious from the definition of the rank distribution in Definition \ref{DEF:RANK_DIST_DOM}. 
\begin{lemma}
\label{lem:rank_dominance_first_down_convex_combination_non_dominance} 
Let $y,y^1,\ldots,y^K \in \Delta(X)$ be assignments such that 
\begin{equation}
	y = \sum_{k = 1}^K y^k \cdot \alpha_k 
\end{equation}
for the same coefficients $\alpha_1, \ldots, \alpha_K$, and let there be $\text{rank}_k,k=1,\ldots,K$ such that for all $k = 1,\ldots,K$ and all $\text{rank}' < \text{rank}_k$ we have
\begin{equation}
	d^{x^k}_{\text{rank}'} = d^{y^k}_{\text{rank}'}.
\label{eq:rank_component_difference_equal}
\end{equation}
Furthermore, if $\text{rank}_k \leq m$, then
\begin{equation}
	d^{x^k}_{\text{rank}_k} > d^{y^k}_{\text{rank}_k}
\label{eq:rank_component_difference_strict}
\end{equation}
(otherwise, $d^{x^k} = d^{y^k}$).

Then $y$ does not even weakly rank dominate $x$. 
\end{lemma}
\begin{proof}[Proof of Lemma \ref{lem:rank_dominance_first_down_convex_combination_non_dominance}:] 
Let $\text{rank}_{\min} = \{\text{rank}_k| k = 1,\ldots,K\}$ be the lowest rank (i.e., the best choice) at which inequality (\ref{eq:rank_component_difference_strict}) holds strictly, and let $k_{\min} \in \{1,\ldots,K+1\}$ be an index for which this is the case. 
Then for all $\text{rank}' < \text{rank}_{\min}$ and all $k=1,\ldots,K$ we have 
\begin{equation}
	d^{x^k}_{\text{rank}'} = d^{y^k}_{\text{rank}'}, 
\end{equation}
so that by Lemma \ref{lem:rank_distribution_of_convex_combinations} 
\begin{equation}
	d^{x}_{\text{rank}'} = d^{y}_{\text{rank}'}. 
\label{eq:rank_comparison_at_ranks_equal}
\end{equation}
In words, the rank distributions of $x$ and $y$ coincides for all ranks before $\text{rank}_{\min}$. 
Furthermore, 
\begin{equation}
	d^{x^k}_{\text{rank}_{\min}} \geq d^{y^k}_{\text{rank}_{\min}}
\end{equation}
for all $k \neq k_{\min}$, and 
\begin{equation}
	d^{x^{k_{\min}}}_{\text{rank}_{\min}} > d^{y^{k_{\min}}}_{\text{rank}_{\min}}.
\end{equation}
Thus, by Lemma \ref{lem:rank_distribution_of_convex_combinations} and the fact that $\alpha_{k_{\min}} > 0$, 
\begin{eqnarray}
	d^{x}_{\text{rank}_{\min}} & = & \sum_{k=1}^K d^{x^k}_{\text{rank}_{\min}} \cdot \alpha_k \\
		& = & d^{x^{k_{\min}}}_{\text{rank}_{\min}} \cdot \alpha_{k_{\min}}  + \sum_{k=1, k \neq k_{\min}}^K d^{x^k}_{\text{rank}_{\min}} \cdot \alpha_k \\
		& > & d^{y^{k_{\min}}}_{\text{rank}_{\min}} \cdot \alpha_{k_{\min}}  + \sum_{k=1, k \neq k_{\min}}^K d^{x^k}_{\text{rank}_{\min}} \cdot \alpha_k \\
		& \geq & \sum_{k=1}^K d^{y^k}_{\text{rank}_{\min}} \cdot \alpha_k = d^{y}_{\text{rank}_{\min}}.
\label{eq:rank_comparison_at_ranks_strict}
\end{eqnarray}
\end{proof}
We now proceed to prove the Theorem in three steps. 

\medskip
\textbf{Step 1 (for any fixed single priority profile):} 
First, we show that for any fixed single priority profile $\pp \in \Pi^M$ and any preference profile $ P$, the assignment $y^{\pp} = \DA^\pp( P)$ never strictly rank dominates the assignment $x^{\pp} = \BM^\pp( P)$ at $ P$. 
In fact, we show something stronger, namely that the conditions of Lemma \ref{lem:rank_dominance_first_down_convex_combination_non_dominance} are satisfied for $x^{\pp}$ and $y^{\pp}$, i.e., either $d^{x^{\pp}} = d^{y^{\pp}}$, or there exists some $r \in \{1,\ldots,m\}$ such that $d^{x^{\pp}}_r > d^{y^{\pp}}_r$, and $d^{x^{\pp}}_{r'} = d^{y^{\pp}}_{r'}$ for all $r' < r$.

In the proof, we consider the slightly larger domain, where schools can have zero capacity. 
Under $\DA^\pp$, including additional empty schools does not make a difference for the resulting assignment. 
Under the $\BM^\pp$, it is easy to see that the assignments can be decomposed into two parts:
\begin{enumerate}
	\setlength{\itemsep}{0pt}
	\item Run the first round of the mechanism, in which a set of students $N_1$ receives their first-choice schools. 
	\item Remove the students $N_1$ from $N$, and also remove these students from all priority orders in the priority profile $\pp$.
		Reduce the capacities of the schools they received by the number of students who received each school. 
		Change the preference orders of all \emph{remaining} students by moving their first choice to the end of their ranking. 
		Then, run the mechanism again on the reduced problem (which may include schools of capacity zero). 
\end{enumerate}
In the final assignment resulting from $\BM^\pp$, the students $N_1$ will receive their first choices, and the other students will receive the schools they got in the reduced setting. 
\begin{claim} 
\label{claim:DA_no_more_first_choices_than_BM} 
$\DA^\pp( P)$ assigns a weakly lower number of first choices than $\BM^\pp( P)$.
\end{claim}
The claim is obvious from the observation that $\BM^\pp$ maximizes the number of assigned first choices. 
\begin{claim} 
\label{claim:same_rank_DA_same_assignment} 
If $\DA^\pp( P)$ assigns the same number of first choices as $\BM^\pp( P)$, then the sets of students who get their first choices under both mechanism coincide. 
\end{claim}
\begin{proof} 
By assumption, $d^{x^{\pp}}_1 = d^{y^{\pp}}_1$. 
Suppose towards contradiction that there exists some student $i \in N$, who receives her first-choice school $j\in M$ under $y^{\pp}$ but not under $x^{\pp}$. 
That means that $j$ was exhausted in the first round by other students, all of whom must have had higher priority than $i$ (according to $\pp$). 
These students as well as $i$ would also apply to $j$ in the first round of $\DA^\pp$. 
But since $j$ was already exhausted by the other students, $i$ will also be rejected from $j$ in the first round of $\DA^\pp$, a contradiction. 
\end{proof} 
Observer that under $\DA^\pp$ students can only obtain their first-choice school in the first round. 
By Claim \ref{claim:same_rank_DA_same_assignment}, if $d^{x^{\pp}}_1 = d^{y^{\pp}}_1$, then $\BM^\pp$ and $\DA^\pp$ assign the same students to their first-choice schools, and therefore, none of the students who received their first-choice school under $\DA^\pp$ (tentatively in the first round) was rejected in any subsequent round. 
Thus, we can also decompose the assignment from $\DA^\pp$ into two parts (as before for $\BM^\pp$):
\begin{enumerate}
	\setlength{\itemsep}{0pt}
	\item The assignment from the first round. 
	\item The assignments from applying the mechanism to the reduced and altered setting. 
\end{enumerate}
We can now apply Claim \ref{claim:same_rank_DA_same_assignment} inductively to the reduced settings to show that $d^{x^{\pp}}_r = d^{y^{\pp}}_r$ implies that the same students also got their $r^{\text{th}}$ choice under both mechanisms. 
Since $d^{x^{\pp}}_r < d^{y^{\pp}}_r$ is impossible by Claim \ref{claim:DA_no_more_first_choices_than_BM}, we get that either the assignments from both mechanism coincide entirely, or $d^{x^{\pp}}_r > d^{y^{\pp}}_r$ for some $r\in \{1,\ldots,m\}$, i.e., the Boston mechanism assigns strictly more $r^{\text{th}}$ choices than Deferred Acceptance. 

\medskip
\textbf{Step 2 (for any single priority distributions):} 
For any single priority distribution $\PP$, a single priority profile $\pp$ is drawn at random according to $\PP$. 
By construction
\begin{equation}
	x = \BM^\PP( P) = \sum_{\pp} \BM^\pp( P) \cdot \PP[\pp],
\end{equation}
and
\begin{equation}
	y = \DA^\PP( P) = \sum_{\pp} \DA^\pp( P) \cdot \PP[\pp],
\end{equation}
i.e., both $x$ and $y$ can we written as convex combinations of assignments $x^{\pp} = \BM^\pp( P)$ and $y^{\pp} = \DA^\pp( P)$ respectively, with the same coefficients $\alpha_{\pp} = \PP[\pp]$. 
By Step 1, each pair $x^{\pp},y^{\pp}$ has the property that 
$d^{x^{\pp}}_{r'} = d^{y^{\pp}}_{r'}$ for $r' < r \leq m$ and $d^{x^{\pp}}_{r} > d^{y^{\pp}}_{r}$ (or $d^{x^{\pp}} = d^{y^{\pp}}$). 
Thus, by Lemma \ref{lem:rank_dominance_first_down_convex_combination_non_dominance}, $\DA^\PP( P)$ never strictly rank dominates $\BM^\PP( P)$.

\medskip
\textbf{Step 3 (for any priority distribution):} 
\citet{Harless2015ImmediateAcceptancePlus} showed that Claim \ref{claim:same_rank_DA_same_assignment} also holds for multiple priority profiles. 
We can apply the same reasoning as in Step 2 to obtain comparable rank dominance of $\BM^\PP$ over $\DA^\PP$ for any priority distribution $\PP$.

This concludes the proof of Theorem \ref{THM:BM_DOM_DA}.}

\newcommand{\ExNbmDaIncomparable}{
Consider a setting with three students $N=\{1,2,3\}$, three schools $M=\{a,b,c\}$ with a single seat each, and the preference profile $ P = (P_1,P_2,P_3)$ with
\begin{equation*}
	P_1~:~a\succ \ldots, 
	\ \ \ \ P_2~:~a \succ b \succ c, 
	\ \ \ \ P_3: b \succ c \succ a. 
\end{equation*}
For the single priority order $ \pi$ with $\pi_j~:~1 \succ 2 \succ 3$, $\BMpi$ assigns 
$a$ to 1, $b$ to 3, and $c$ to 2. 
The resulting cumulative rank distribution is 
$c^{\BMpi( P)} = (2,2,3)$. 
For the same single priority order, $\DApi$ assigns
$a$ to 1, $b$ to 2, and $c$ to 3, which leads to the cumulative rank distribution 
$c^{\DApi( P)} = (1,3,3)$. 
Observe that neither assignment rank dominates the other. 
Thus, $\BMpi$ and $\DApi$ are not comparable by rank dominance at this preference profile.%
}

\newcommand{\ThmDaNotDomAbmLimitStatement}{%
Let $(N^n,M^n, q^n)_{n\geq 1}$ be a sequence of settings such that
\begin{enumerate}
\setlength{\itemsep}{0pt}
	\item \label{THM:ABM_DOM_DA_LIMIT:SC} either $M^n = M$ and $n = \sum_{j\in M} q_j^n$ for all $n$, and $\min_{j \in M} q_j \overset{(n \rightarrow \infty)}{\longrightarrow} \infty$, 
	\item \label{THM:ABM_DOM_DA_LIMIT:HA} or $\left|M^n\right| = n$ and $q_j^n = 1$ for all $n$ and all $j \in M^n$, 
\end{enumerate}
then
\begin{equation}
		\frac{\# \{  P~:~\DAU( P) \text{ weakly rank dominates } \ABMU( P) \text{ at }  P \}}{\# \{  P \text{ preference profile} \}} \overset{(n \rightarrow \infty)}{\longrightarrow} 0.
\end{equation}%
}
\newcommand{\TheoremSchoolChoiceLimitProof}{
An assignment $x$ is \emph{first-choice-maximizing} at preference profile $ P$ if it can be represented as a lottery over deterministic assignments that give the maximum number of first choices, i.e.,
\begin{equation}
	d^x_1 = \sum_{i \in N} x_{i,j} \mathds{1}_{r_i(j) = 1} = \max_{y \in \mathcal{X}} d^y_1.
\end{equation}
Since any ex-post efficient assignment is supported by a serial dictatorship, $\DA^\UU$ puts positive probability on \emph{all} ex-post efficient, deterministic assignments.
In contrast, $\ABM^\UU$ assigns positive probabilities to only \emph{some} ex-post efficient, deterministic assignments.
In particular, $\ABM^\UU$ is first choice maximizing, i.e., it gives no probability to any assignment that does not yield the maximum possible number of first choices.
Consequently, if at some preference profile $ P$ there exists at least one ex-post efficient, deterministic assignment that is not first-choice-maximizing, then $\DA^\UU$ will assign strictly less first choices than $\ABM^\UU$. 
At these preference profiles, $\DA^\UU$ is guaranteed not to rank dominate $\ABM^\UU$ (even weakly).

Using this observation, we can now prove the following Claim \ref{claim:sc_markets_limit} which in turn yields the result.
\begin{claim} \label{claim:sc_markets_limit} 
For any fixed number of schools $m \geq 3$ and any $\epsilon > 0$, there exists $q_{\min} \in \mathds{N}$, such that for any capacities $q_1, \ldots, q_m$ with $q_j \geq q_{\min}$ for all $j \in M$ and $n = \sum_{j \in M} q_j$ students, and for $ P$ chosen uniformly at random from $\mathcal{P}^n$, the probability that $\DA^\UU( P)$ is first-choice-maximizing is smaller than $\epsilon$.
\end{claim}
For a given preference profile $ P \in \mathcal{P}^N$, the \emph{first choice profile} $\mathbf{k}^{ P} = (k_j^{ P})_{j\in M}$ is the vector of non-negative integers, where $k_j^{ P}$ represents the number of students whose first choice is $j$.
For a fixed setting, i.e., the triple $(N,M,\mathbf{q})$, we consider a uniform distribution on the space of preference profiles $\mathcal{P}^n$. 
As the preference profile $ P$ is held fixed, we suppress the index and simply write $k_j$.
We say that a school $j \in M$ is
\begin{itemize}
	\setlength{\itemsep}{0pt}
	\item \emph{un-demanded} if $k_j = 0$,
	\item \emph{under-demanded} if $k_j \in \{1,\ldots,q_j-1\}$,
	\item \emph{exhaustively demanded} if $k_j = q_j$,
	\item and \emph{over-demanded} if $k_j > q_j$.
\end{itemize}
For any first choice profile $\mathbf{k}^{ P}$, one of the following cases must hold:
\begin{enumerate}
	\setlength{\itemsep}{0pt}
	\item \label{case:one_undemanded} There is at least one un-demanded school.
	\item \label{case:no_overdemanded} All schools are exhaustively demanded.
	\item \label{case:one_over_one_exhausted} No school is un-demanded, at least one school is over-demanded, and at least one other school is exhaustively demanded.
	\item \label{case:two_overdemanded} No school is un-demanded, but at least two schools are over-demanded.
	\item \label{case:exactly_one_over} There is exactly one over-demanded school, and all other schools are under-demanded.
\end{enumerate}
We will show that for fixed $m$ and increasing minimum capacity, the probabilities for cases (\ref{case:one_undemanded}) and (\ref{case:no_overdemanded}) become arbitrarily small. 
We will further show that in cases (\ref{case:one_over_one_exhausted}), (\ref{case:two_overdemanded}), and (\ref{case:exactly_one_over}), the probabilities that $\DA^\UU$ assigns the maximum number of first choices become arbitrarily small.
\begin{enumerate}
	\item The probability that under a randomly chosen preference profile at least one school is un-demanded is upper-bounded by
		\begin{equation}
			\frac{\binom{m}{1}(m-1)^n}{m^n} = m \left(\frac{m-1}{m}\right)^n,
		\end{equation}
		which converges to 0 as $n = \sum_{j \in M} q_j \geq m q_{\min}$ becomes large (where $m$ is fixed).
	\item Let $\tilde{q} = \frac{n}{m}$. 
		Without loss of generality, $\tilde{q}$ can be chosen as a natural number (otherwise, we increase the capacity of the school with least capacity until $n$ is divisible by $m$). 
		The probability that under a randomly chosen preference profile all schools are exhaustively demanded is
		\begin{equation}
			\frac{\binom{n}{q_1,\ldots,q_m}}{m^n} \leq \frac{\binom{n}{\tilde{q},\ldots,\tilde{q}}}{m^n} \lesssim \frac{(m\tilde{q})^{m\tilde{q}}}{(\tilde{q})^{m\tilde{q}} m^{m\tilde{q}}}\sqrt{\frac{m\tilde{q}}{\tilde{q}^m}} = \sqrt{\frac{m}{\tilde{q}^{m-1}}},
		\end{equation}
		which converges to 0 as $\tilde{q} \geq q_{\min}$ becomes large (where $m$ is fixed).
	\item Suppose that $\DA^\UU$ is first choice maximizing.
		If one school $a$ is over-demanded and another school $b$ is exhaustively demanded, then no student with first choice $a$ can have $b$ as second choice. 
		Otherwise, there exists an order of the student such that a student with first choice $a$ will get $b$. 
		In that case, $b$ is not assigned entirely to students with first choice $b$, and hence, the assignment can not maximize the number of first choices. 
		Thus, the probability that the $k_a$ students who have first choice $a$ all have a second choice different from $b$ (conditional on the first choice profile) is
		\begin{equation}
			\left(\frac{m-2}{m-1}\right)^{k_1} < \left(\frac{m-2}{m-1}\right)^{q_1} \leq \left(\frac{m-2}{m-1}\right)^{q_{\min}}.
		\end{equation}
		This becomes arbitrarily small for increasing $q_{\min}$. Thus, the probability that the maximum number of first choices is assigned by $\DA^\UU$, conditional on case (\ref{case:one_over_one_exhausted}) becomes small.
	\item This case is analogous to (\ref{case:one_over_one_exhausted}).
	\item Suppose that for some preference profile consistent with case (\ref{case:exactly_one_over}), $\DA^\UU$ assigns the maximum number of first choices. 
		Let $a$ be the school that is over-demanded and let $j_2,\ldots,j_m$ be the under-demanded schools. 
		Then the maximum number of first choices is assigned if and only if
		\begin{itemize}
			\setlength{\itemsep}{0pt}
			\item $q_{a}$ students with first choice $a$ receive $a$, and
			\item all students with first choices $j_2,\ldots,j_m$ receive their respective first choice.
		\end{itemize}
		If $\DA^\UU$ maximizes the number of first choices, then for any ordering of the students, the maximum number of first choices must be assigned, i.e., the two conditions are true. 
		If the students with first choice $a$ get to pick before all other students, then they exhaust $a$ and get at most $q_{j}-k_{j}$ of the schools $j\neq a$: otherwise, if they got more than $q_j-k_j$ of school $j$, then some student with first choice $j$ would get a worse choice, which violates first choice maximization.
		
		After any $q_a$ of the $k_a$ students with first choice $a$ consume school $a$, there are $k_a-q_a$ students left which will consume other schools.
		Since $n = \sum_{j\in M} q_j = \sum_{j \in M} k_j$, we get that
		\begin{equation}
			k_a-q_a = n-\sum_{j \neq a} k_j - (n - \sum_{j \neq a} q_j) = \sum_{j \neq a} q_j - k_j.
		\end{equation}
		Therefore, the second choice profile of these $k_a-q_a$ students must be $(l_2,\ldots,l_m)$, where $l_r = q_{j_r} - k_{j_r} \geq 1$. 
		In addition, some student $i'$ who consumed $a$ has second choice $j'$, and some student $i''$ with first choice $a$ gets its second choice $j'' \neq j'$. 
		If we exchange the place of $i'$ and $i''$ in the ordering, $i''$ will get $a$ and $i'$ will get $j'$. 
		But then $q_{j'}-k_{j'}+1$ students with first choice $a$ get their second choice $j'$. 
		Therefore, when the students with first choice $j'$ get to pick their schools, there are only $k_{j'}-1$ copies of $j'$ left, which is not sufficient. 
		Thus, we have constructed an ordering of the students under which the number of assigned first choices is not maximized. 
		This implies that for any preference profile with first choice profile satisfying case (\ref{case:exactly_one_over}), $\DA^\UU$ does not assign the maximum number of first choices.
\end{enumerate}
Combining the arguments for all cases, we can find $q_{\min}$ sufficiently high, such that we can estimate the probability that $\DA^\UU$ maximizes first choices (\emph{$\DA^\UU$ mfc.}) by
\begin{eqnarray}
\mathds{Q}[\DA^\UU\text{ mfc.}] & = & \mathds{Q}[\DA^\UU\text{ mfc.}|\text{(\ref{case:one_undemanded})}] \mathds{Q}[\text{(\ref{case:one_undemanded})}] \\
	& & +	\mathds{Q}[\DA^\UU\text{ mfc.}|\text{(\ref{case:no_overdemanded})}] \mathds{Q}[\text{(\ref{case:no_overdemanded})}]  \\
	& & + \mathds{Q}[\DA^\UU\text{ mfc.}|\text{(\ref{case:one_over_one_exhausted})}] \mathds{Q}[\text{(\ref{case:one_over_one_exhausted})}] \\
	& & + \mathds{Q}[\DA^\UU\text{ mfc.}|\text{(\ref{case:two_overdemanded})}] \mathds{Q}[\text{(\ref{case:two_overdemanded})}] \\
	& & +	\mathds{Q}[\DA^\UU\text{ mfc.}|\text{(\ref{case:exactly_one_over})}] \mathds{Q}[\text{(\ref{case:exactly_one_over})}]  \\
	& \leq & \mathds{Q}[\text{(\ref{case:one_undemanded})}] + \mathds{Q}[\text{(\ref{case:no_overdemanded})}] + \mathds{Q}[\DA^\UU\text{ mfc.}|\text{(\ref{case:one_over_one_exhausted})}] \\
	& & + \mathds{Q}[\DA^\UU\text{ mfc.}|\text{(\ref{case:two_overdemanded})}] +	 \mathds{Q}[\DA^\UU\text{ mfc.}|\text{(\ref{case:exactly_one_over})}] \\
	& \leq  & \frac{\epsilon}{4} + \frac{\epsilon}{4} + \frac{\epsilon}{4} + \frac{\epsilon}{4} + 0 = \epsilon.
\end{eqnarray}
Here, $\mathds{Q}$ is the probability measure induced by the random selection of a preference profile. 
}

\newcommand{\TheoremHouseAssignmentLimitProof}{
As for the proof of case (\ref{THM:ABM_DOM_DA_LIMIT:SC}), we establish that $\DA^\UU$ is almost never first choice maximizing at a randomly selected preference profile.
\begin{claim}
\label{claim:house_markets_limit}
For any $\epsilon > 0$, there exists $n \in \mathds{N}$, such that for any setting $(N,M,\mathbf{q})$ with $\#M = \#N \geq n $ and $q_j =1 $ for all $j \in M$, and for $ P$ chosen uniformly at random, the probability that $\DA^\UU( P)$ is first-choice-maximizing is smaller than $\epsilon$.
\end{claim}

Recall that for a fixed preference profile $ P$, $\DA^\UU$ is first-choice maximizing only if all ex-post efficient assignments are first-choice maximizing. 
We will introduce \emph{no overlap}, a necessary condition on the preference profile that ensures that $\DA^\UU$ assigns the maximum number of first choices. 
Conversely, if a preference profile violates \emph{no overlap}, $\DA^\UU$ will not assign the maximum number of first choices. 
To establish Claim \ref{claim:house_markets_limit}, we show that the share of preference profiles that exhibit no overlap vanishes as $n$ becomes large.

The proof requires some more formal definitions: 
for convenience, we will enumerate the set $M$ of schools by the integers $\{1,\ldots,n\}$. 
As in the proof of case (\ref{THM:ABM_DOM_DA_LIMIT:SC}), $\mathbf{k}^{ P} = (k^{ P}_1,\ldots,k^{ P}_n)$ is called the \emph{first choice profile} of the type profile $ P$, where $k^{ P}_j$ is the number of students whose first choice is school $j$. 
To reduce notation, we suppress the superscript $ P$. 
For some first choice profile $\mathbf{k}$ and school $j$ we define the following indicators:
\begin{equation}
w_{\mathbf{k}}(j)
	:= \left\{ \begin{array}{ll} 1, & \text{ if }k_j \geq 1 \\ 0, & \text{else}  \end{array} \right.
	\text{ and }
o_{\mathbf{k}}(j)
	:= \left\{ \begin{array}{ll} 1, & \text{ if }k_j \geq 2 \\ 0, & \text{else}.  \end{array} \right.
\end{equation}
$w$ indicates whether $j$ is \emph{demanded}, i.e., it is the first choice of at least one student, and $o$ indicates whether $j$ is \emph{over-demanded}, i.e., it is the first choice of more than one student. 
Further, we define
\begin{equation}
W_{\mathbf{k}} := \sum_{j \in M} w_{\mathbf{k}}(j), \;\;\; O_{\mathbf{k}} := \sum_{j \in M} o_{\mathbf{k}}(j), \;\;\; C_{\mathbf{k}} = \sum_{j \in M} k_j \cdot o_{\mathbf{k}}(j).
\end{equation} 
$W_{\mathbf{k}}$ is the number of schools that are demanded by at least one student, $O_{\mathbf{k}}$ is the number of over-demanded schools, and $C_{\mathbf{k}}$ is the number of students competing for over-demanded schools.
Finally, a preference profile $ P$ \emph{exhibits overlap} if there exists a student $i\in N$ with first choice $j_1$ and second choice $j_2$, such that $o_{\mathbf{k}^{ P}}(j_1) = 1 $ and $w_{\mathbf{k}^{ P}}(j_2) = 1$, i.e., student $i$'s first choice is over-demanded and its second choice is demanded as a first choice by at least one other student. 
As an example consider a setting where three students have preferences
\begin{equation}  P_1 : a \succ \ldots, \ \ \ \succ_2 : a \succ b \succ \ldots, \ \ \ \succ_3 : b \succ \ldots.
\label{eq:overlap_example}
\end{equation}
The maximum number of first choices that can be assigned is $2$, e.g., by giving $a$ to 1 and $b$ to $3$. 
But for the priority order $1~\pi~2~\pi~3$, student $1$ will get $a$ and student $2$ will get $b$. 
Then student $3$ cannot take $b$, and consequently $\DA^\UU$ will not assign the maximum number of first choices. 
If a preference profile exhibits overlap, a situation as in (\ref{eq:overlap_example}) will arise for some priority order, and therefore, $\DA^\UU$ will not assign the maximum number of first choices. 
Conversely, no overlap in $ P$ is a necessary condition for $\DA^\UU( P)$ to assign the maximum number of first choices. 
We will show in the following that the share of preference profiles exhibiting no overlap becomes small for increasing $n$.

Consider a uniform distribution (denoted $\mathds{Q}$) on the preference profiles, i.e., all students draw their preference order independently and uniformly at random from the space of all possible preference orders. 
Then the statement that \emph{the share of preference profiles exhibiting no overlap becomes small} is equivalent to the statement that the \emph{probability of selecting a preference profile with no overlap converges to 0}. 
The proof of the following Claim \ref{claim:QNoOverlapToZero} is technical and requires involved combinatorial and asymptotic arguments.
\begin{claim} 
\label{claim:QNoOverlapToZero}
$\mathds{Q}[  P\text{ no overlap}] \rightarrow 0$ for $n \rightarrow \infty$.
\end{claim}
\begin{proof}[Proof of Claim \ref{claim:QNoOverlapToZero}]
Using conditional probability, we can write the probability that a preference profile is without overlap as
\begin{equation}
\mathds{Q}[ P\text{ no overlap}] = \sum_{\mathbf{k}} \mathds{Q}[\mathbf{k} = \mathbf{k}^{ P}] \cdot \mathds{Q}[ P\text{ no overlap}\ |\ \mathbf{k}=\mathbf{k}^{ P}].
\label{eq:no_overlap_cond_prob}
\end{equation}
The number of preference profiles that have first choice profile $\mathbf{k} = (k_1,\ldots,k_n)$ is proportional to the number of ways to distribute $n$ unique balls (students) across $n$ urns (first choices), such that $k_j$ balls end up in urn $j$. 
Thus,
\begin{equation}
\mathds{Q}[\mathbf{k} = \mathbf{k}^{ P}] = \frac{\binom{n}{k_1, \ldots, k_n}(n-1)!^n}{(n!)^n} = \frac{\binom{n}{k_1, \ldots, k_n}}{n^n}.
\label{eq:number_types_with_fcp_k}
\end{equation}
In order to ensure no overlap, a student with an over-demanded first choice cannot have as her second choice a school that is the first choice of any other student.
Students whose first choice is not over-demanded can have any school (except for their own first choice) as second choice. 
Thus, given a first choice profile $\mathbf{k}$, the conditional probability of no overlap is
\begin{eqnarray}
\mathds{Q}[ P \text{ no overlap} \ |\ \mathbf{k}=\mathbf{k}^{ P}] 
	& = & \prod_{j \in M} \left( \left(1-o_{\mathbf{k}}(j)\right) + o_{\mathbf{k}}(j)\left(\frac{n-W_{\mathbf{k}}}{n-1}\right)^{k_j} \right) 	\\
	& = & \left(\frac{n-W_{\mathbf{k}}}{n-1}\right)^{\sum_{j \in M} k_j \cdot o_{\mathbf{k}}(j)}=  \left(\frac{n-W_{\mathbf{k}}}{n-1}\right)^{C_{\mathbf{k}}} \\
	& = & \left(\frac{C_{\mathbf{k}}-O_{\mathbf{k}}}{n-1}\right)^{C_{\mathbf{k}}},
\end{eqnarray}
where the last equality holds, since $ n-W_{\mathbf{k}} = n - (n-C_{\mathbf{k}} + O_{\mathbf{k}}) = C_{\mathbf{k}} - O_{\mathbf{k}} $. 
Thus, the probability of no overlap can be determined as
\begin{equation}
\mathds{Q}[ P\text{ no overlap}] = \frac{1}{n^n} \sum_{\mathbf{k}} \binom{n}{k_1, \ldots, k_n} \left(\frac{C_{\mathbf{k}} - O_{\mathbf{k}}}{n-1}\right)^{C_{\mathbf{k}}}.
\label{eq:prob_no_overlap_1}
\end{equation}
$C_{\mathbf{k}}$ is either $0$ or $\geq 2$, since a single student cannot be in competition. 
If no students compete ($C_{\mathbf{k}} = 0$), all must have different first choices. 
Thus, for $\mathbf{k} = (1,\ldots,1)$, the term in the sum in (\ref{eq:prob_no_overlap_1}) is
\begin{equation}
\binom{n}{k_1, \ldots, k_n} \left(\frac{C_{\mathbf{k}} - O_{\mathbf{k}}}{n-1}\right)^{C_{\mathbf{k}}} = \binom{n}{1, \ldots, 1} \cdot 1 = n!.
\label{eq:no_comp_summand}
\end{equation}
Using this and sorting the terms for summation by $c$ for $C_{\mathbf{k}}$ and $o$ for $O_{\mathbf{k}}$, we get
\begin{equation}
\mathds{Q}[ P\text{ no overlap}] = \frac{1}{n^n} \left[n! + \sum_{c=2}^n\sum_{o=1}^{\left\lfloor \frac{c}{2} \right\rfloor} \left(\frac{c - o}{n-1}\right)^{c} \sum_{\mathbf{k}: C_{\mathbf{k}} = c, O_{\mathbf{k}} = o}  \binom{n}{k_1, \ldots, k_n} \right].
\label{eq:prob_no_overlap_2}
\end{equation}
Consider the inner sum
\begin{equation}
\sum_{\mathbf{k}: C_{\mathbf{k}} = c, O_{\mathbf{k}} = o}  \binom{n}{k_1, \ldots, k_n}
\label{eq:constraint_sum_1}
\end{equation}
in (\ref{eq:prob_no_overlap_2}): with a first choice profile $\mathbf{k}$ that satisfies $C_{\mathbf{k}} = c$ and $O_{\mathbf{k}} = o$ there are exactly $o$ over-demanded schools (i.e., schools $j$ with $k_j \geq 2$), $n-c$ singly-demanded schools (with $k_j = 1$), and $c-o$ un-demanded schools (with $k_j = 0$). 
Therefore,
\begin{eqnarray}
\sum_{\mathbf{k}: C_{\mathbf{k}} = c, O_{\mathbf{k}} = o}  \binom{n}{k_1, \ldots, k_n} & = & \binom{n}{c-o} \sum_{\mathbf{k'} = (k_1', \ldots, k'_{n-c+o}): C_{\mathbf{k'}} = c, O_{\mathbf{k'}} = o}  \binom{n}{k'_1, \ldots, k'_{n-c+o}} \\
	& = & \binom{n}{c-o} \binom{n-c+o}{n-c} \frac{n!}{c!} \sum_{\mathbf{k''} = (k_1'', \ldots, k''_{o}): k''_j \geq 2} \binom{c}{k''_1, \ldots, k''_o}. \label{eq:constraint_sum_2}
\end{eqnarray}
The first equality holds because we simply choose $c-o$ of the $n$ schools to be un-demanded, and
\begin{equation}
\binom{n}{k_1, \ldots, k_{r-1},0,k_{r+1},\ldots,k_m} = \binom{n}{k_1, \ldots, k_{r-1},k_{r+1},\ldots,k_m}.
\label{eq:binomi_reduce}
\end{equation}
The second equality holds because we select the $n-c$ singly-demanded schools from the remaining $n-c+o$ schools as well as the $n-c$ students to demand them.
The sum (\ref{eq:constraint_sum_2}) is equal to the number of ways to distribute $c$ unique balls to $o$ unique urns such that each urn contains at least 2 balls. 
This in turn is equal to
\begin{equation}
o! \left\{\hspace{-0.05in}\left\{ \begin{array}{c} c \\ o \end{array} \right\}\hspace{-0.05in}\right\},
\label{eq:stirling_1}
\end{equation}
where $\left\{\hspace{-0.03in}\left\{ : \right\}\hspace{-0.03in}\right\}$ denotes the 2-associated Stirling number of the second kind. 
This number represents the number of ways to partition $c$ unique balls such that each partition contains at least 2 balls. 
The factor $o!$ in (\ref{eq:stirling_1}) is included to make the partitions unique. 
$\left\{\hspace{-0.03in}\left\{ : \right\}\hspace{-0.03in}\right\}$ is upper-bounded by $\left\{ : \right\}$, the Stirling number of the second kind, which represents the number of ways to partition $c$ unique balls such that no partition is empty. 
Furthermore, the Stirling number of the second kind has the upper bound
\begin{equation}
\left\{ \begin{array}{c} c \\ o \end{array} \right\} \leq \binom{c}{o} o^{c-o}.
\label{eq:stirling_bound_1}
\end{equation}
Thus, the sum in (\ref{eq:constraint_sum_2}) can be upper-bounded by
\begin{equation}
\sum_{\mathbf{k''} = (k_1'', \ldots, k''_{o}): k''_j \geq 2} \binom{c}{k''_1, \ldots, k''_o} \leq o! \binom{c}{o}o^{c-o}.
\label{eq:eq:constraint_sum_3}
\end{equation}
Combining all the previous observations, we can estimate the probability $\mathds{Q}[ P\text{ no overlap}]$ from (\ref{eq:prob_no_overlap_2}) by
\begin{equation}
\mathds{Q}[ P\text{ no overlap}] \leq \frac{1}{n^n} \left[n! + \sum_{c=2}^n\sum_{o=1}^{\left\lfloor \frac{c}{2} \right\rfloor} \left(\frac{c - o}{n-1}\right)^{c} \binom{n}{c-o}\binom{n-c+o}{n-c}\binom{c}{o}\frac{n!o!}{c!}o^{c-o} \right].
\label{eq:prob_no_overlap_3}
\end{equation}
The Stirling approximation yields
\begin{equation}
\sqrt{2\pi} e^{\frac{1}{12n+1}} \leq \frac{n!}{\sqrt{n}\left(\frac{n}{e}\right)^n} \leq \sqrt{2\pi} e^{\frac{1}{12n}},
\label{eq:stirling_approx}
\end{equation}
and therefore $n! \approx \left(\frac{n}{e}\right)^n \sqrt{n}$ up to a constant factor. Using this, we observe that the first term in (\ref{eq:prob_no_overlap_3}) converges to 0 as $n$ increases, i.e.,
\begin{equation}
\frac{n!}{n^n} \approx \frac{\sqrt{n}}{e^n} \rightarrow 0 \text{ for }n \rightarrow \infty.
\label{eq:n_fact_vanish}
\end{equation}
Now we need to estimate the double sum in (\ref{eq:prob_no_overlap_3}):
\begin{eqnarray}
& & \frac{1}{n^n} \sum_{c=2}^n\sum_{o=1}^{\left\lfloor \frac{c}{2} \right\rfloor} \left(\frac{c - o}{n-1}\right)^{c} \binom{n}{c-o}\binom{n-c+o}{n-c}\binom{c}{o}\frac{n!o!}{c!}o^{c-o} \\
& = & \frac{n!}{n^n} \sum_{c=2}^n\sum_{o=1}^{\left\lfloor \frac{c}{2} \right\rfloor} \frac{n!}{(c-o)!(n-c+o)!} \cdot \frac{(n-c+o)!}{o!(n-c)!} \cdot \frac{c!}{o!(c-o)!} \cdot \frac{o!}{c!} \cdot \frac{(c-o)^c o^{c-o}}{(n-1)^c} \\
& = & \frac{n!}{n^n} \sum_{c=2}^n\sum_{o=1}^{\left\lfloor \frac{c}{2} \right\rfloor} \binom{n}{c} \binom{c}{o} \left(\frac{n}{n-1}\right)^c \frac{1}{n^c}  \cdot \frac{(c-o)^c o^{c-o}}{(c-o)!} \\
& \lesssim & \left[\sqrt{n}\left(\frac{n}{n-1}\right)^{n-1}\left(\frac{n-1}{n}\right)\right]\frac{1}{e^n} \sum_{c=2}^n\sum_{o=1}^{\left\lfloor \frac{c}{2} \right\rfloor} \binom{n}{c} \binom{c}{o} \frac{1}{n^c}  \cdot \frac{(c-o)^c o^{c-o}}{(c-o)^{c-o}}e^{c-o} \\
& \leq & \left[e\sqrt{n}\right]\frac{1}{e^n} \sum_{c=2}^n\sum_{o=1}^{\left\lfloor \frac{c}{2} \right\rfloor} \binom{n}{c} \binom{c}{o} \frac{1}{n^c}  \cdot (c-o)^o o^{c-o} e^{c-o}, \label{eq:estimate_1}
\end{eqnarray}
where we use that $\left(1+\frac{x}{n}\right)^n \leq e^x$. Using the binomial theorem and the fact that the function
$o \mapsto (c-o)^o o^{c-o}$ is maximized by $o = \frac{c}{2}$, we can further estimate (\ref{eq:estimate_1}) by
\begin{eqnarray}
\left[e\sqrt{n}\right]\frac{1}{e^n} \sum_{c=2}^n \binom{n}{c}\left(\frac{ec}{2n}\right)^c  \sum_{o=1}^{\left\lfloor \frac{c}{2} \right\rfloor} \binom{c}{o} \left(\frac{1}{e}\right)^o
& \leq & \left[e\sqrt{n}\right]\frac{1}{e^n} \sum_{c=2}^n \binom{n}{c}\left(\frac{ec}{2n}\right)^c  \left(1 + \frac{1}{e}\right)^c \\
& \leq & \left[e\sqrt{n}\right]\frac{1}{e^n} \sum_{c=2}^n \binom{n}{c}\left(\alpha \cdot \frac{c}{n}\right)^c  \label{eq:estimate_2}
\end{eqnarray}
with $\alpha = \frac{(1+e)}{2}$.
To estimate the sum in (\ref{eq:estimate_2}), we first consider even $n$ and note the following:
\begin{itemize}
	\item $\alpha = \frac{(1+e)}{2} \approx 1.85914\ldots < e$, and therefore, the last term of the sum for $c=n$ can be ignored as $\binom{n}{n}\left(\frac{\alpha}{e}\right)^n \rightarrow 0$ for $n \rightarrow \infty$.
	\item $\binom{n}{c} = \binom{n}{n-c}$, and therefore, both terms $\binom{n}{c}\left(\alpha \frac{c}{n}\right)^c$ and $\binom{n}{c}\left(\alpha \frac{n-c}{n}\right)^{n-c}$ have the same binomial coefficient in the sum.
	\item The idea is to estimate the sum of both terms by an exponential function of the form $c \mapsto e^{m c + b}$, where $m$ and $b$ depend only on $n$ and $\alpha$.
	\item Indeed, the log of the sum, the function $c \mapsto \log\left( \left(\alpha\frac{c}{n}\right)^c + \left(\alpha\frac{n-c}{n}\right)^{n-c} \right)$, is strictly convex and on the interval $\left[1,\frac{n}{2}\right]$ it is upper-bounded by the linear function
	\begin{equation}
		f(c) = \left(\frac{\log(4)}{n}-\log(2\alpha)\right) c + n \log(\alpha).
	\label{eq:linear_bound_1}
	\end{equation}
	\item Thus,
	\begin{equation}
		\binom{n}{c}\left(\alpha\frac{c}{n}\right)^c + \binom{n}{n-c}\left(\alpha\frac{n-c}{n}\right)^{n-c} \leq \binom{n}{c} e^{f(c)}.
	\label{eq:linear_bound_2}
	\end{equation}
\end{itemize}
We can bound (\ref{eq:estimate_2}) by
\begin{eqnarray}
\left[e\sqrt{n}\right]\frac{1}{e^n} \sum_{c=1}^{ \frac{n}{2} } \binom{n}{c} e^{f(c)} & = & \left[e\sqrt{n}\right]\frac{1}{e^n} \sum_{c=1}^{\frac{n}{2} } \binom{n}{c} 4^{\frac{c}{n}} \alpha^n \left(\frac{1}{2\alpha}\right)^c \\
& \leq & \left[4e\sqrt{n}  \right] \frac{\alpha^n \left(1+ \frac{1}{2\alpha}\right)^n}{e^n} \\
& = & \left[4e\sqrt{n}  \right] \left(\frac{\frac{1}{2} + \alpha}{e}\right)^n  \approx \left[4e\sqrt{n}  \right]\left(\frac{2.35914\ldots}{e}\right)^n.
\end{eqnarray}
Since $2.35914 < e$, the exponential convergence of the last term dominates the divergence of the first terms, which is of the order $\sqrt{n}$, and the expression converges to $0$.

For odd $n$ the argument is essentially the same, except that we need to also consider the central term (for $c = \frac{n}{2}+1$) separately.
\begin{eqnarray}
\binom{n}{\frac{n}{2}+1} \left(\alpha \frac{\frac{n}{2}+1}{2}\right)^{\frac{n}{2}+1} & \leq & 2^n  \left( \sqrt{\alpha \left(\frac{1}{2} + \frac{1}{n}\right)}\right)^n \cdot \alpha \left(\frac{1}{2} + \frac{1}{n}\right) \\
& = & \left( \sqrt{\alpha \left(2 + \frac{4}{n}\right)}\right)^n \cdot \alpha \left(\frac{1}{2} + \frac{1}{n}\right).
\end{eqnarray}
With $\sqrt{\alpha \left(2 + \frac{4}{n}\right)} \approx 1.92828\ldots < e$, the result follows for odd $n$ as well. \end{proof}}
\newcommand{\ThmDaNotDomAbmLimitProof}{%
First, we show case (\ref{THM:ABM_DOM_DA_LIMIT:SC}) of Theorem \ref{THM:ABM_DOM_DA_LIMIT}:
\TheoremSchoolChoiceLimitProof

\medskip
Next, we show case (\ref{THM:ABM_DOM_DA_LIMIT:HA}) of Theorem \ref{THM:ABM_DOM_DA_LIMIT}:
\TheoremHouseAssignmentLimitProof

This completes the proof of Theorem \ref{THM:ABM_DOM_DA_LIMIT}.%
}

\newcommand{\PlotsNbmPiDaPiRankDomComparison}{
\begin{tikzpicture}[scale=1]
\begin{axis}[
    ybar stacked,
		xlabel={{Number of students $n$}},
    legend style={at={(0.95,0.80)},
      anchor=east,legend columns=1},
    ylabel={Share of profiles (in \%)},
    xtick={50, 100, 200, 300, 400, 500, 600, 700, 800, 900, 1000},
    xticklabels={$50$, $100$, $200$, $300$, $400$, $500$, $600$, $700$, $800$, $900$, $\numprint{1000}$},
		width=14cm, 
		height=5.8cm,
		legend cell align=left,
    ]
\addplot+[ybar,Brown!80] plot coordinates {
	(50,22.02)	(100,18.6)	(150,16.16)	(200,14.14)	(250,13.4)	(300,13.13)	(350,11.55)	(400,11.24)	(450,10.79)	(500,10.43)	(550,9.62)	(600,8.66)	(650,9.64)	(700,8.96)	(750,8.67)	(800,9.29)	(850,8.02)	(900,8.13)	(950,7.8)	(1000,7.95)
};
\addplot+[ybar,Brown!50] plot coordinates {
	(50,2.23)	(100,0.62)	(150,0.25)	(200,0.12)	(250,0.03)	(300,0)	(350,0.02)	(400,0.01)	(450,0)	(500,0)	(550,0)	(600,0)	(650,0)	(700,0)	(750,0)	(800,0)	(850,0)	(900,0)	(950,0)	(1000,0)
};
\legend{\strut {$\BMpi$ strictly rank dominates $\DApi$}, \strut {$\BMpi$ and $\DApi$ have same rank distribution}}
\end{axis}
\end{tikzpicture}
}

\newcommand{\PlotsAbmPiDaPiRankDomComparison}{
\begin{tikzpicture}[scale=1]
\begin{axis}[
    ybar stacked,
		xlabel={{Number of students $n$}},
    legend style={at={(0.95,0.80)},
      anchor=east,legend columns=1},
    ylabel={Share of profiles (in \%)},
    xtick={50, 100, 200, 300, 400, 500, 600, 700, 800, 900, 1000},
    xticklabels={$50$, $100$, $200$, $300$, $400$, $500$, $600$, $700$, $800$, $900$, $\numprint{1000}$},
		width=14cm, 
		height=5.8cm,
		legend cell align=left,
    ]
\addplot+[ybar,Black] plot coordinates {
	(50,0.1)	(100,0.04)	(150,0)	(200,0.01)	(250,0.01)	(300,0)	(350,0)	(400,0)	(450,0)	(500,0)	(550,0)	(600,0)	(650,0)	(700,0)	(750,0)	(800,0)	(850,0)	(900,0)	(950,0)	(1000,0)
};
\addplot+[ybar,Brown!80] plot coordinates {
	(50,19.67)	(100,16.87)	(150,14.58)	(200,13.05)	(250,12.11)	(300,11.91)	(350,10.39)	(400,10.3)	(450,9.95)	(500,8.94)	(550,8.2)	(600,8.1)	(650,8.25)	(700,7.91)	(750,7.54)	(800,7.57)	(850,7.08)	(900,7.37)	(950,6.94)	(1000,7.26)
};
\addplot+[ybar,Brown!50] plot coordinates {
	(50,3.57)	(100,1.01)	(150,0.51)	(200,0.18)	(250,0.11)	(300,0.03)	(350,0.06)	(400,0.01)	(450,0.03)	(500,0.03)	(550,0.01)	(600,0)	(650,0)	(700,0.01)	(750,0)	(800,0)	(850,0)	(900,0.01)	(950,0)	(1000,0)
};
\legend{\strut {$\DApi$ strictly rank dominates $\ABMpi$}, \strut {$\ABMpi$ strictly rank dominates $\DApi$}, \strut {$\ABMpi$ and $\DApi$ have same rank distribution}}
\end{axis}
\end{tikzpicture}
}

\newcommand{\PlotsNbmPiAbmPiRankDomComparison}{
\begin{tikzpicture}[scale=1]
\begin{axis}[
    ybar stacked,
		xlabel={{Number of students $n$}},
    legend style={at={(0.99,0.85)},
      anchor=east,legend columns=1},
    ylabel={Share of profiles (in \%)},
    xtick={50, 100, 200, 300, 400, 500, 600, 700, 800, 900, 1000},
    xticklabels={$50$, $100$, $200$, $300$, $400$, $500$, $600$, $700$, $800$, $900$, $\numprint{1000}$},
		width=14cm, 
		height=5.8cm,
		legend cell align=left,
    ]
\addplot+[ybar,Black] plot coordinates {
	(50,0.55)	(100,0.47)	(150,0.46)	(200,0.51)	(250,0.41)	(300,0.49)	(350,0.26)	(400,0.36)	(450,0.46)	(500,0.33)	(550,0.25)	(600,0.18)	(650,0.15)	(700,0.22)	(750,0.19)	(800,0.17)	(850,0.15)	(900,0.16)	(950,0.12)	(1000,0.11)
};
\addplot+[ybar,Brown!80] plot coordinates {
	(50,25.74)	(100,26.78)	(150,27.7)	(200,26.77)	(250,26.36)	(300,26.57)	(350,25.78)	(400,25.59)	(450,24.18)	(500,23.78)	(550,22.99)	(600,22.51)	(650,22.45)	(700,22.39)	(750,21.85)	(800,22.13)	(850,21.25)	(900,21.14)	(950,20.55)	(1000,20.69)	
};
\addplot+[ybar,Brown!50] plot coordinates {
	(50,32.53)	(100,21.04)	(150,14.19)	(200,11.3)	(250,8.7)	(300,7.15)	(350,5.56)	(400,4.8)	(450,4.41)	(500,3.83)	(550,3.15)	(600,2.92)	(650,2.41)	(700,2.5)	(750,1.99)	(800,1.8)	(850,1.77)	(900,1.58)	(950,1.29)	(1000,1.29)	
};
\legend{\strut {$\ABMpi$ strictly rank dominates $\BMpi$}, \strut {$\BMpi$ strictly rank dominates $\ABMpi$}, \strut {$\BMpi$ and $\ABMpi$ have same rank distribution}}
\end{axis}
\end{tikzpicture}
}

\newcommand{\PlotsTopKFactorEffects}{
\begin{tikzpicture}[scale=1]
\begin{groupplot}[group style={group name=my plots, group size = 3 by 1}]
	  \nextgroupplot[
			mark size=1.5,
			xlabel={Number of students $n$}, 
			ylabel={Rank transition $K$}, 
			ymin=0,
			width=6cm,
			height=5cm,
			xtick={100,400,700,1000},
			xticklabels={$100$,$400$,$700$,$\numprint{1000}$},
			]
			\addplot[color=BrickRed,mark=x] coordinates { 
				(100,2)
				(200,2)
				(300,2)
				(400,2)
				(500,2)
				(600,2)
				(700,2)
				(800,2)
				(900,2)
				(1000,2)
			};
			\addplot[color=blue,mark=square] coordinates { 
				(100,3)
				(200,3)
				(300,3)
				(400,3)
				(500,3)
				(600,3)
				(700,3)
				(800,3)
				(900,3)
				(1000,3)
			};
			\addplot[color=OliveGreen,mark=o] coordinates { 
				(100,5)
				(200,5)
				(300,5)
				(400,5)
				(500,5)
				(600,5)
				(700,5)
				(800,5)
				(900,5)
				(1000,5)
			};
		\nextgroupplot[ 
			mark size=1.5,
			xlabel={{Number of schools $m$}}, 
			ymin=0,
			width=6cm,
			height=5cm,
			xtick={10,40,70,100},
			legend style={at={(0.5,1.1)}, anchor=south,legend columns=-1},
			] 
			\addplot[color=BrickRed,mark=x] coordinates { 
				(10,2)
				(15,2)
				(20,2)
				(25,2)
				(30,2)
				(35,3)
				(40,3)
				(45,3)
				(50,3)
				(55,3)
				(60,3)
				(65,4)
				(70,4)
				(75,4)
				(80,4)
				(85,4)
				(90,4)
				(95,5)
				(100,5)
			};
			\addplot[color=blue,mark=square] coordinates { 
				(10,3)
				(15,4)
				(20,4)
				(25,5)
				(30,5)
				(35,6)
				(40,6)
				(45,7)
				(50,7)
				(55,8)
				(60,8)
				(65,8)
				(70,9)
				(75,9)
				(80,10)
				(85,10)
				(90,10)
				(95,11)
				(100,11)
			};
			\addplot[color=OliveGreen,mark=o] coordinates { 
				(10,5)
				(15,7)
				(20,8)
				(25,10)
				(30,11)
				(35,12)
				(40,13)
				(45,15)
				(50,16)
				(55,16)
				(60,18)
				(65,19)
				(70,19)
				(75,20)
				(80,22)
				(85,22)
				(90,23)
				(95,24)
				(100,24)
			}; 
			\legend{{$K(\ABM,\DA)$},{$K(\BM,\DA)$},{$K(\BM,\ABM)$}};
		\nextgroupplot[ 
			mark size=1.5,
			xlabel={Correlation $\alpha$}, 
			xtick={0,.25,.5,.75,.95},
			ymin=0,
			width=6cm,
			height=5cm,
			] 
			\addplot[color=BrickRed,mark=x] coordinates { 
				(0,2)
				(0.05,2)
				(0.1,2)
				(0.15,2)
				(0.2,2)
				(0.25,2)
				(0.3,2)
				(0.35,4)
				(0.4,5)
				(0.45,5)
				(0.5,7)
				(0.55,7)
				(0.6,8)
				(0.65,8)
				(0.7,8)
				(0.75,8)
				(0.8,8)
				(0.85,8)
				(0.9,8)
				(0.95,8)
			};
			\addplot[color=blue,mark=square] coordinates { 
				(0,3)
				(0.05,3)
				(0.1,3)
				(0.15,3)
				(0.2,3)
				(0.25,3)
				(0.3,4)
				(0.35,5)
				(0.4,6)
				(0.45,7)
				(0.5,7)
				(0.55,8)
				(0.6,8)
				(0.65,8)
				(0.7,8)
				(0.75,8)
				(0.8,8)
				(0.85,8)
				(0.9,8)
				(0.95,8)
			};
			\addplot[color=OliveGreen,mark=o] coordinates { 
				(0,5) (0.05,5) (0.1,5) (0.15,4) (0.2,5) (0.25,5)
				(0.3,5)
				(0.35,6)
				(0.4,7)
				(0.45,7)
				(0.5,8)
				(0.55,8)
				(0.6,8)
				(0.65,8)
				(0.7,8)
				(0.75,8)
				(0.8,8)
				(0.85,8)
				(0.9,8) (0.95,8)
		}; 
\end{groupplot}
\node[below = 1.1cm of my plots c1r1.south] {(a)};
\node[below = 1.1cm of my plots c2r1.south] {(b)};
\node[below = 1.1cm of my plots c3r1.south] {(c)};
\end{tikzpicture}
}

\newcommand{\PlotsMexRankDistTenSingle}{
\begin{tikzpicture}[scale=1]
\begin{axis}[
			mark size=1.5,
			xlabel={Rank}, 
			ylabel={Number of students}, 
			width=14cm,
			height=8cm,
			ytick scale label code/.code={},
			ytick={100000,150000,200000,250000},
			yticklabels={100k,150k,200k,250k},
			xtick={1,5,10,15,20},
			legend style={at={(.95,.1)}, anchor=south east},
			]
			\addplot[color=BrickRed,mark=x] coordinates { 
				(1,136994) (2,151427.5) (3,165379.6) (4,178029.4) (5,188569.5) (6,196192.7) (7,202402.8) (8,207069.5) (9,210793.4) (10,213325.4) (11,214827) (12,215961.1) (13,216837.9) (14,217485.4) (15,217958.6) (16,218313) (17,218592.5) (18,218826.3) (19,219027.9) (20,219175.1) 
			};
			\addlegendentry{\BM}
			\addplot[color=blue,mark=square] coordinates { 
				(1,136994) (2,146674.8) (3,157901.2) (4,169294.3) (5,180246.8) (6,189788.9) (7,197706.5) (8,204081.2) (9,209210.9) (10,213151.7) (11,215756.1) (12,217519.4) (13,218784.9) (14,219723.2) (15,220398.3) (16,220835) (17,221148) (18,221399.6) (19,221591.3) (20,221724.6) 
			};
			\addlegendentry{\ABM}
			\addplot[color=OliveGreen,mark=o] coordinates { 
				(1,100166) (2,126247.1) (3,146222.9) (4,163200.1) (5,177830.1) (6,189900) (7,199541) (8,207102.1) (9,212996.2) (10,217437.1) (11,220266.4) (12,222148) (13,223447.6) (14,224386.4) (15,225059.3) (16,225487.3) (17,225788.8) (18,226017.6) (19,226184) (20,226299.8) 
			};
			\addlegendentry{\DA}
			\draw[densely dashed] ({rel axis cs:0,.1} -| {axis cs:5,0}) -- ({rel axis cs:0,1} -| {axis cs:5,0});
			\draw[densely dashed] ({rel axis cs:0,.3} -| {axis cs:7,0}) -- ({rel axis cs:0,1} -| {axis cs:7,0});
			\draw[densely dashed] ({rel axis cs:0,.5} -| {axis cs:10,0}) -- ({rel axis cs:0,1} -| {axis cs:10,0});
			\node at ({rel axis cs:0,.1} -| {axis cs:5,0}) [anchor=west] {$K(\ABM,\DA) = 5$};
			\node at ({rel axis cs:0,.3} -| {axis cs:7,0}) [anchor=west] {$K(\BM,\DA) = 7$};
			\node at ({rel axis cs:0,.5} -| {axis cs:10,0}) [anchor=west] {$K(\BM,\ABM) = 10$};
\end{axis}
\end{tikzpicture}
}

\newcommand{\PlotsMexRankDistElevenSingle}{
\begin{tikzpicture}[scale=1]
\begin{axis}[
			mark size=1.5,
			xlabel={Rank}, 
			ylabel={Number of students}, 
			width=14cm,
			height=8cm,
			ytick scale label code/.code={},
			ytick={100000,150000,200000,250000},
			yticklabels={100k,150k,200k,250k},
			xtick={1,5,10,15,20},
			legend style={at={(.95,.1)}, anchor=south east},
			]
			\addplot[color=BrickRed,mark=x] coordinates { 
				(1,134372) (2,148684.1) (3,162081.6) (4,174515) (5,185855) (6,193877.8) (7,200128.2) (8,205710.9) (9,209735.8) (10,212729.1) (11,214495.9) (12,215733.9) (13,216723.6) (14,217431.4) (15,217982.9) (16,218414.9) (17,218764) (18,219012.4) (19,219211.7) (20,219410.1) 
			};
			\addlegendentry{\BM}
			\addplot[color=blue,mark=square] coordinates { 
				(1,134372) (2,144211.9) (3,155646.2) (4,167109.4) (5,178158.2) (6,187950.4) (7,196239.9) (8,203001.5) (9,208331) (10,212544.6) (11,215296.2) (12,217242.9) (13,218617.3) (14,219647.9) (15,220401.3) (16,220901.8) (17,221261.6) (18,221522.2) (19,221717.8) (20,221883.5) 
			};
			\addlegendentry{\ABM}
			\addplot[color=OliveGreen,mark=o] coordinates { 
				(1,99265) (2,125045.9) (3,145140.5) (4,161945.6) (5,176424.8) (6,188493.6) (7,198359.8) (8,206189.3) (9,212283.8) (10,216917.7) (11,219950.8) (12,221972.7) (13,223383.9) (14,224410.8) (15,225142.4) (16,225609.8) (17,225929.7) (18,226160.8) (19,226332.6) (20,226462.2) 
			};
			\addlegendentry{\DA}
			\draw[densely dashed] ({rel axis cs:0,.1} -| {axis cs:5,0}) -- ({rel axis cs:0,1} -| {axis cs:5,0});
			\draw[densely dashed] ({rel axis cs:0,.3} -| {axis cs:7,0}) -- ({rel axis cs:0,1} -| {axis cs:7,0});
			\draw[densely dashed] ({rel axis cs:0,.5} -| {axis cs:10,0}) -- ({rel axis cs:0,1} -| {axis cs:10,0});
			\node at ({rel axis cs:0,.1} -| {axis cs:5,0}) [anchor=west] {$K(\ABM,\DA) = 5$};
			\node at ({rel axis cs:0,.3} -| {axis cs:7,0}) [anchor=west] {$K(\BM,\DA) = 7$};
			\node at ({rel axis cs:0,.5} -| {axis cs:10,0}) [anchor=west] {$K(\BM,\ABM) = 10$};
\end{axis}
\end{tikzpicture}
}

\newcommand{\PlotsMexRankDistTwelveSingle}{
\begin{tikzpicture}[scale=1]
\begin{axis}[
			mark size=1.5,
			xlabel={Rank}, 
			ylabel={Number of students}, 
			width=14cm,
			height=8cm,
			ytick scale label code/.code={},
			ytick={100000,150000,200000,250000},
			yticklabels={100k,150k,200k,250k},
			xtick={1,5,10,15,20},
			legend style={at={(.95,.1)}, anchor=south east},
			]
			\addplot[color=BrickRed,mark=x] coordinates { 
				(1,133119) (2,148280.4) (3,160845.7) (4,173228.7) (5,184568.8) (6,193697.9) (7,201130.3) (8,206786.3) (9,211262.7) (10,214482.3) (11,216653.3) (12,218176.8) (13,219260) (14,220057.8) (15,220711.1) (16,221163.1) (17,221529.2) (18,221825.2) (19,222100.4) (20,222298.6) 
			};
			\addlegendentry{\BM}
			\addplot[color=blue,mark=square] coordinates { 
				(1,133119) (2,143078.7) (3,154423.3) (4,166198.3) (5,177794.6) (6,188287.9) (7,197056.4) (8,204128.6) (9,209929.3) (10,214405) (11,217439.2) (12,219540.4) (13,221016.4) (14,222171.6) (15,223010.2) (16,223564.3) (17,223948.3) (18,224271.2) (19,224509.4) (20,224686.7) 
 			};
			\addlegendentry{\ABM}
			\addplot[color=OliveGreen,mark=o] coordinates { 
				(1,98807.42) (2,124761.97) (3,144993.23) (4,162058.6) (5,176932.79) (6,189528.27) (7,199778.26) (8,207840.81) (9,214249.52) (10,219108.69) (11,222329.21) (12,224511.7) (13,226021.05) (14,227134.03) (15,227929.01) (16,228463.6) (17,228830.71) (18,229102.18) (19,229306.61) (20,229444.46)
			};
			\addlegendentry{\DA}
			\draw[densely dashed] ({rel axis cs:0,.1} -| {axis cs:5,0}) -- ({rel axis cs:0,1} -| {axis cs:5,0});
			\draw[densely dashed] ({rel axis cs:0,.3} -| {axis cs:7,0}) -- ({rel axis cs:0,1} -| {axis cs:7,0});
			\draw[densely dashed] ({rel axis cs:0,.5} -| {axis cs:10,0}) -- ({rel axis cs:0,1} -| {axis cs:10,0});
			\node at ({rel axis cs:0,.1} -| {axis cs:5,0}) [anchor=west] {$K(\ABM,\DA) = 5$};
			\node at ({rel axis cs:0,.3} -| {axis cs:7,0}) [anchor=west] {$K(\BM,\DA) = 7$};
			\node at ({rel axis cs:0,.5} -| {axis cs:10,0}) [anchor=west] {$K(\BM,\ABM) = 10$};
\end{axis}
\end{tikzpicture}
}

\newcommand{\PlotsMexRankDistThirteenSingle}{
\begin{tikzpicture}[scale=1]
\begin{axis}[
			mark size=1.5,
			xlabel={Rank}, 
			ylabel={Number of students}, 
			width=14cm,
			height=8cm,
			ytick scale label code/.code={},
			ytick={100000,150000,200000,250000},
			yticklabels={100k,150k,200k,250k},
			xtick={1,5,10,15,20},
			legend style={at={(.95,.1)}, anchor=south east},
			]
			\addplot[color=BrickRed,mark=x] coordinates { 
				(1,134915) (2,149785.7) (3,163154.5) (4,175394.2) (5,187447.4) (6,197570.2) (7,204871.9) (8,210769.5) (9,216237.9) (10,219930.6) (11,222389.4) (12,224182.5) (13,225574.1) (14,226552.8) (15,227357.6) (16,227924.5) (17,228319.5) (18,228660.5) (19,228935.4) (20,229172.9) 
			};
			\addlegendentry{\BM}
			\addplot[color=blue,mark=square] coordinates { 
				(1,134915) (2,144999.2) (3,156702) (4,168846) (5,180854.2) (6,191890.8) (7,201216.2) (8,208891.8) (9,215161.6) (10,220198.8) (11,223570.5) (12,225942.9) (13,227631.4) (14,228923.8) (15,229897.8) (16,230551.7) (17,231016.1) (18,231379.8) (19,231634.4) (20,231847.7) 
			};
			\addlegendentry{\ABM}
			\addplot[color=OliveGreen,mark=o] coordinates { 
				(1,99029.71) (2,126109.94) (3,147122.58) (4,164693.55) (5,180082.82) (6,193311.8) (7,204157.62) (8,212853.4) (9,219797.05) (10,225196.76) (11,228749.66) (12,231153.38) (13,232808.16) (14,234044.06) (15,234958.32) (16,235554.8) (17,235974.11) (18,236292.64) (19,236513.59) (20,236675.55) 
			};
			\addlegendentry{\DA}
			\draw[densely dashed] ({rel axis cs:0,.1} -| {axis cs:5,0}) -- ({rel axis cs:0,1} -| {axis cs:5,0});
			\draw[densely dashed] ({rel axis cs:0,.3} -| {axis cs:7,0}) -- ({rel axis cs:0,1} -| {axis cs:7,0});
			\draw[densely dashed] ({rel axis cs:0,.5} -| {axis cs:9,0}) -- ({rel axis cs:0,1} -| {axis cs:9,0});
			\node at ({rel axis cs:0,.1} -| {axis cs:5,0}) [anchor=west] {$K(\ABM,\DA) = 5$};
			\node at ({rel axis cs:0,.3} -| {axis cs:7,0}) [anchor=west] {$K(\BM,\DA) = 7$};
			\node at ({rel axis cs:0,.5} -| {axis cs:9,0}) [anchor=west] {$K(\BM,\ABM) = 9$};
\end{axis}
\end{tikzpicture}
}

\newcommand{\PlotsMexRankDistFourteenSingle}{
\begin{tikzpicture}[scale=1]
\begin{axis}[
			mark size=1.5,
			xlabel={Rank}, 
			ylabel={Number of students}, 
			width=14cm,
			height=8cm,
			ytick scale label code/.code={},
			ytick={100000,150000,200000,250000},
			yticklabels={100k,150k,200k,250k},
			xtick={1,5,10,15,20},
			legend style={at={(.95,.1)}, anchor=south east},
			]
			\addplot[color=BrickRed,mark=x] coordinates { 
				(1,134545)	(2,148797.23)	(3,162391.84)	(4,174541.6)	(5,186875.76)	(6,197002.38)	(7,204495.36)	(8,211055.93)	(9,216595.56)	(10,220217.16)	(11,222615.21)	(12,224378.7)	(13,225755.4)	(14,226904.6)	(15,227808.29)	(16,228414.34)	(17,228841.29)	(18,229240.51)	(19,229582.41)	(20,229867.03)
			};
			\addlegendentry{\BM}
			\addplot[color=blue,mark=square] coordinates { 
				(1,134545)	(2,144365)	(3,155975.51)	(4,168029.81)	(5,180076.23)	(6,191157.87)	(7,200592.39)	(8,208551.83)	(9,215025.56)	(10,220391.1)	(11,223873.01)	(12,226374.11)	(13,228162.21)	(14,229507.34)	(15,230526.42)	(16,231218.43)	(17,231736.92)	(18,232149.89)	(19,232461.21)	(20,232702.31)
			};
			\addlegendentry{\ABM}
			\addplot[color=OliveGreen,mark=o] coordinates { 
				(1,97102.78)	(2,124049.78)	(3,144940.16)	(4,162682.76)	(5,178351.88)	(6,191869.13)	(7,203087.73)	(8,212327.01)	(9,219674.27)	(10,225485.94)	(11,229239.14)	(12,231809.71)	(13,233610.6)	(14,234942.76)	(15,235919.05)	(16,236587.74)	(17,237071.12)	(18,237431.75)	(19,237706.2)	(20,237909.04)
			};
			\addlegendentry{\DA}
			\draw[densely dashed] ({rel axis cs:0,.1} -| {axis cs:5,0}) -- ({rel axis cs:0,1} -| {axis cs:5,0});
			\draw[densely dashed] ({rel axis cs:0,.3} -| {axis cs:7,0}) -- ({rel axis cs:0,1} -| {axis cs:7,0});
			\draw[densely dashed] ({rel axis cs:0,.5} -| {axis cs:9,0}) -- ({rel axis cs:0,1} -| {axis cs:9,0});
			\node at ({rel axis cs:0,.1} -| {axis cs:5,0}) [anchor=west] {$K(\ABM,\DA) = 5$};
			\node at ({rel axis cs:0,.3} -| {axis cs:7,0}) [anchor=west] {$K(\BM,\DA) = 7$};
			\node at ({rel axis cs:0,.5} -| {axis cs:9,0}) [anchor=west] {$K(\BM,\ABM) = 9$};
\end{axis}
\end{tikzpicture}
}

\newcommand{\PlotsMexRankDistFourteenSingleMain}{
\begin{tikzpicture}[scale=1]
\begin{axis}[
			mark size=1.5,
			xlabel={Rank}, 
			ylabel={Number of students}, 
			width=14cm,
			height=7cm,
			ytick scale label code/.code={},
			ytick={100000,150000,200000,250000},
			yticklabels={100k,150k,200k,250k},
			xtick={1,5,10,15,20},
			legend style={at={(.95,.1)}, anchor=south east},
			]
			\addplot[color=BrickRed,mark=x] coordinates { 
				(1,134545)	(2,148797.23)	(3,162391.84)	(4,174541.6)	(5,186875.76)	(6,197002.38)	(7,204495.36)	(8,211055.93)	(9,216595.56)	(10,220217.16)	(11,222615.21)	(12,224378.7)	(13,225755.4)	(14,226904.6)	(15,227808.29)	(16,228414.34)	(17,228841.29)	(18,229240.51)	(19,229582.41)	(20,229867.03)
			};
			\addlegendentry{\BM}
			\addplot[color=blue,mark=square] coordinates { 
				(1,134545)	(2,144365)	(3,155975.51)	(4,168029.81)	(5,180076.23)	(6,191157.87)	(7,200592.39)	(8,208551.83)	(9,215025.56)	(10,220391.1)	(11,223873.01)	(12,226374.11)	(13,228162.21)	(14,229507.34)	(15,230526.42)	(16,231218.43)	(17,231736.92)	(18,232149.89)	(19,232461.21)	(20,232702.31)
			};
			\addlegendentry{\ABM}
			\addplot[color=OliveGreen,mark=o] coordinates { 
				(1,97102.78)	(2,124049.78)	(3,144940.16)	(4,162682.76)	(5,178351.88)	(6,191869.13)	(7,203087.73)	(8,212327.01)	(9,219674.27)	(10,225485.94)	(11,229239.14)	(12,231809.71)	(13,233610.6)	(14,234942.76)	(15,235919.05)	(16,236587.74)	(17,237071.12)	(18,237431.75)	(19,237706.2)	(20,237909.04)
			};
			\addlegendentry{\DA}
			\draw[densely dashed] ({rel axis cs:0,.1} -| {axis cs:5,0}) -- ({rel axis cs:0,1} -| {axis cs:5,0});
			\draw[densely dashed] ({rel axis cs:0,.3} -| {axis cs:7,0}) -- ({rel axis cs:0,1} -| {axis cs:7,0});
			\draw[densely dashed] ({rel axis cs:0,.5} -| {axis cs:9,0}) -- ({rel axis cs:0,1} -| {axis cs:9,0});
			\node at ({rel axis cs:0,.1} -| {axis cs:5,0}) [anchor=west] {$K(\ABM,\DA) = 5$};
			\node at ({rel axis cs:0,.3} -| {axis cs:7,0}) [anchor=west] {$K(\BM,\DA) = 7$};
			\node at ({rel axis cs:0,.5} -| {axis cs:9,0}) [anchor=west] {$K(\BM,\ABM) = 9$};
\end{axis}
\end{tikzpicture}
}

\newcommand{\PlotsMexRankDistTenMultiple}{
\begin{tikzpicture}[scale=1]
\begin{axis}[
			mark size=1.5,
			xlabel={Rank}, 
			ylabel={Number of students}, 
			width=14cm,
			height=8cm,
			ytick scale label code/.code={},
			ytick={50000,100000,150000,200000,250000},
			yticklabels={50k,100k,150k,200k,250k},
			xtick={1,5,10,15,20},
			legend style={at={(.95,.1)}, anchor=south east},
			]
			\addplot[color=BrickRed,mark=x] coordinates { 
				(1,136994) (2,151429.5) (3,165371.3) (4,178015.7) (5,188549) (6,196164) (7,202374.4) (8,207032.3) (9,210754.4) (10,213291.9) (11,214796.6) (12,215931.6) (13,216813.6) (14,217466.1) (15,217940.2) (16,218294.2) (17,218572.5) (18,218805.8) (19,219007.2) (20,219155.1) 
			};
			\addlegendentry{\BM}
			\addplot[color=blue,mark=square] coordinates { 
				(1,136994) (2,147095.9) (3,158519.1) (4,169983.5) (5,180893.6) (6,190337.4) (7,198102.2) (8,204314.6) (9,209307.2) (10,213139.3) (11,215665.5) (12,217367.2) (13,218591.3) (14,219497.9) (15,220152) (16,220576.1) (17,220875.6) (18,221120.9) (19,221303.2) (20,221436.4) 
			};
			\addlegendentry{\ABM}
			\addplot[color=OliveGreen,mark=o] coordinates { 
				(1,47055.91) (2,83266.26) (3,113636.35) (4,139830.65) (5,162027.56) (6,180009.79) (7,194253.68) (8,205157.19) (9,213346.18) (10,219386.84) (11,222942.24) (12,225239.87) (13,226735.74) (14,227736.04) (15,228401.7) (16,228807.63) (17,229074.06) (18,229257.81) (19,229383.09) (20,229468.11) 
			};
			\addlegendentry{\DA}
			\draw[densely dashed] ({rel axis cs:0,.1} -| {axis cs:7,0}) -- ({rel axis cs:0,1} -| {axis cs:7,0});
			\draw[densely dashed] ({rel axis cs:0,.3} -| {axis cs:8,0}) -- ({rel axis cs:0,1} -| {axis cs:8,0});
			\draw[densely dashed] ({rel axis cs:0,.5} -| {axis cs:10,0}) -- ({rel axis cs:0,1} -| {axis cs:10,0});
			\node at ({rel axis cs:0,.1} -| {axis cs:7,0}) [anchor=west] {$K(\ABM,\DA) = 7$};
			\node at ({rel axis cs:0,.3} -| {axis cs:8,0}) [anchor=west] {$K(\BM,\DA) = 8$};
			\node at ({rel axis cs:0,.5} -| {axis cs:10,0}) [anchor=west] {$K(\BM,\ABM) = 10$};
\end{axis}
\end{tikzpicture}
}

\newcommand{\PlotsMexRankDistElevenMultiple}{
\begin{tikzpicture}[scale=1]
\begin{axis}[
			mark size=1.5,
			xlabel={Rank}, 
			ylabel={Number of students}, 
			width=14cm,
			height=8cm,
			ytick scale label code/.code={},
			ytick={50000,100000,150000,200000,250000},
			yticklabels={50k,100k,150k,200k,250k},
			xtick={1,5,10,15,20},
			legend style={at={(.95,.1)}, anchor=south east},
			]
			\addplot[color=BrickRed,mark=x] coordinates { 
				(1,134372) (2,148683.5) (3,162074.9) (4,174503.2) (5,185840.7) (6,193862.2) (7,200113.6) (8,205693.5) (9,209709.4) (10,212711.9) (11,214474.1) (12,215712.3) (13,216701.3) (14,217410.5) (15,217962.7) (16,218395) (17,218745.2) (18,218993.8) (19,219194) (20,219392.2) 
			};
			\addlegendentry{\BM}
			\addplot[color=blue,mark=square] coordinates { 
				(1,134372) (2,144685.5) (3,156292) (4,167847.2) (5,178840.3) (6,188530.1) (7,196660.3) (8,203265.9) (9,208453.2) (10,212536) (11,215191.5) (12,217074.1) (13,218405.7) (14,219394.4) (15,220124.6) (16,220606.2) (17,220958.1) (18,221209.1) (19,221401) (20,221564.1) 
			};
			\addlegendentry{\ABM}
			\addplot[color=OliveGreen,mark=o] coordinates { 
				(1,47987.56) (2,84289.9) (3,114574.59) (4,140501.89) (5,162362.5) (6,180083.83) (7,194189.39) (8,205068.35) (9,213248.21) (10,219278.96) (11,222925.72) (12,225285.88) (13,226829.16) (14,227869.43) (15,228551.98) (16,228971.07) (17,229238.1) (18,229415.38) (19,229536.95) (20,229621.3) 
			};
			\addlegendentry{\DA}
			\draw[densely dashed] ({rel axis cs:0,.1} -| {axis cs:7,0}) -- ({rel axis cs:0,1} -| {axis cs:7,0});
			\draw[densely dashed] ({rel axis cs:0,.3} -| {axis cs:8,0}) -- ({rel axis cs:0,1} -| {axis cs:8,0});
			\draw[densely dashed] ({rel axis cs:0,.5} -| {axis cs:10,0}) -- ({rel axis cs:0,1} -| {axis cs:10,0});
			\node at ({rel axis cs:0,.1} -| {axis cs:7,0}) [anchor=west] {$K(\ABM,\DA) = 7$};
			\node at ({rel axis cs:0,.3} -| {axis cs:8,0}) [anchor=west] {$K(\BM,\DA) = 8$};
			\node at ({rel axis cs:0,.5} -| {axis cs:10,0}) [anchor=west] {$K(\BM,\ABM) = 10$};
\end{axis}
\end{tikzpicture}
}

\newcommand{\PlotsMexRankDistTwelveMultiple}{
\begin{tikzpicture}[scale=1]
\begin{axis}[
			mark size=1.5,
			xlabel={Rank}, 
			ylabel={Number of students}, 
			width=14cm,
			height=8cm,
			ytick scale label code/.code={},
			ytick={50000,100000,150000,200000,250000},
			yticklabels={50k,100k,150k,200k,250k},
			xtick={1,5,10,15,20},
			legend style={at={(.95,.1)}, anchor=south east},
			]
			\addplot[color=BrickRed,mark=x] coordinates { 
				(1,133119) (2,148283.1) (3,160843.9) (4,173215.5) (5,184556.2) (6,193686.1) (7,201105) (8,206778.1) (9,211244.9) (10,214456.7) (11,216628) (12,218151.9) (13,219234.8) (14,220030.7) (15,220685) (16,221135.4) (17,221503.5) (18,221799.3) (19,222074.1) (20,222271.8) 
			};
			\addlegendentry{\BM}
			\addplot[color=blue,mark=square] coordinates { 
				(1,133119) (2,143581.2) (3,155177.7) (4,167086.5) (5,178636.5) (6,188958.6) (7,197533.8) (8,204394.3) (9,210017.8) (10,214353.7) (11,217303) (12,219346.1) (13,220767.4) (14,221897.9) (15,222714.8) (16,223250.5) (17,223626) (18,223939.6) (19,224180.1) (20,224356.7) 
			};
			\addlegendentry{\ABM}
			\addplot[color=OliveGreen,mark=o] coordinates { 
				(1,50236.79) (2,87023.7) (3,117287.23) (4,143110.23) (5,165000.98) (6,182869.89) (7,197110.2) (8,208086.52) (9,216396.9) (10,222486.87) (11,226235.05) (12,228685.65) (13,230274.5) (14,231355.7) (15,232066.46) (16,232512.68) (17,232802.48) (18,233002.26) (19,233138.55) (20,233230.87) 
			};
			\addlegendentry{\DA}
			\draw[densely dashed] ({rel axis cs:0,.1} -| {axis cs:7,0}) -- ({rel axis cs:0,1} -| {axis cs:7,0});
			\draw[densely dashed] ({rel axis cs:0,.3} -| {axis cs:7,0}) -- ({rel axis cs:0,1} -| {axis cs:7,0});
			\draw[densely dashed] ({rel axis cs:0,.5} -| {axis cs:10,0}) -- ({rel axis cs:0,1} -| {axis cs:10,0});
			\node at ({rel axis cs:0,.1} -| {axis cs:7,0}) [anchor=west] {$K(\ABM,\DA) = 7$};
			\node at ({rel axis cs:0,.3} -| {axis cs:7,0}) [anchor=west] {$K(\BM,\DA) = 7$};
			\node at ({rel axis cs:0,.5} -| {axis cs:10,0}) [anchor=west] {$K(\BM,\ABM) = 10$};
\end{axis}
\end{tikzpicture}
}

\newcommand{\PlotsMexRankDistThirteenMultiple}{
\begin{tikzpicture}[scale=1]
\begin{axis}[
			mark size=1.5,
			xlabel={Rank}, 
			ylabel={Number of students}, 
			width=14cm,
			height=8cm,
			ytick scale label code/.code={},
			ytick={50000,100000,150000,200000,250000},
			yticklabels={50k,100k,150k,200k,250k},
			xtick={1,5,10,15,20},
			legend style={at={(.95,.1)}, anchor=south east},
			]
			\addplot[color=BrickRed,mark=x] coordinates { 
				(1,134915) (2,149785.1) (3,163146.6) (4,175380.4) (5,187437) (6,197555.6) (7,204852.2) (8,210743.8) (9,216210) (10,219896.9) (11,222351) (12,224145.5) (13,225541.4) (14,226525.7) (15,227326.1) (16,227893.4) (17,228289.2) (18,228632.2) (19,228908) (20,229146.4) 
			};
			\addlegendentry{\BM}
			\addplot[color=blue,mark=square] coordinates { 
				(1,134915) (2,145490.3) (3,157389.7) (4,169654.9) (5,181611.6) (6,192497.6) (7,201657.4) (8,209143.3) (9,215271.3) (10,220176.2) (11,223460) (12,225771.3) (13,227410.2) (14,228666.3) (15,229613.5) (16,230251.6) (17,230707.2) (18,231064.2) (19,231321.7) (20,231533.5) 
			};
			\addlegendentry{\ABM}
			\addplot[color=OliveGreen,mark=o] coordinates { 
				(1,56068.26) (2,93136.5) (3,123396.03) (4,149201.21) (5,171235.93) (6,189438.14) (7,204024.54) (8,215387.89) (9,224002.56) (10,230388.98) (11,234312.74) (12,236881.3) (13,238552.81) (14,239695.98) (15,240474.42) (16,240962.86) (17,241288.29) (18,241515.5) (19,241666.71) (20,241771.98) 
			};
			\addlegendentry{\DA}
			\draw[densely dashed] ({rel axis cs:0,.1} -| {axis cs:6,0}) -- ({rel axis cs:0,1} -| {axis cs:6,0});
			\draw[densely dashed] ({rel axis cs:0,.3} -| {axis cs:7,0}) -- ({rel axis cs:0,1} -| {axis cs:7,0});
			\draw[densely dashed] ({rel axis cs:0,.5} -| {axis cs:9,0}) -- ({rel axis cs:0,1} -| {axis cs:9,0});
			\node at ({rel axis cs:0,.1} -| {axis cs:6,0}) [anchor=west] {$K(\ABM,\DA) = 6$};
			\node at ({rel axis cs:0,.3} -| {axis cs:7,0}) [anchor=west] {$K(\BM,\DA) = 7$};
			\node at ({rel axis cs:0,.5} -| {axis cs:9,0}) [anchor=west] {$K(\BM,\ABM) = 9$};
\end{axis}
\end{tikzpicture}
}

\newcommand{\PlotsMexRankDistFourteenMultiple}{
\begin{tikzpicture}[scale=1]
\begin{axis}[
			mark size=1.5,
			xlabel={Rank}, 
			ylabel={Number of students}, 
			width=14cm,
			height=8cm,
			ytick scale label code/.code={},
			ytick={50000,100000,150000,200000,250000},
			yticklabels={50k,100k,150k,200k,250k},
			xtick={1,5,10,15,20},
			legend style={at={(.95,.1)}, anchor=south east},
			]
			\addplot[color=BrickRed,mark=x] coordinates { 
				(1,134545)	(2,148798.31)	(3,162374.31)	(4,174524.36)	(5,186854.71)	(6,196982.97)	(7,204465.35)	(8,211032.45)	(9,216576.26)	(10,220189.78)	(11,222585.52)	(12,224346.16)	(13,225726.02)	(14,226873.89)	(15,227778.84)	(16,228388.41)	(17,228814.52)	(18,229212.45)	(19,229554.41)	(20,229838.71)
			};
			\addlegendentry{\BM}
			\addplot[color=blue,mark=square] coordinates { 
				(1,134545)	(2,144867.9)	(3,156633.07)	(4,168748.26)	(5,180751.41)	(6,191695.11)	(7,200979.6)	(8,208776.95)	(9,215126.48)	(10,220362.83)	(11,223756.76)	(12,226184.14)	(13,227937.63)	(14,229242.67)	(15,230234.8)	(16,230915.71)	(17,231419.12)	(18,231827.64)	(19,232135.28)	(20,232375.99)
};
			\addlegendentry{\ABM}
			\addplot[color=OliveGreen,mark=o] coordinates { 
				(1,53920.11)	(2,90593.71)	(3,120647.35)	(4,146553.63)	(5,168883.03)	(6,187469.25)	(7,202558.12)	(8,214528.98)	(9,223682.25)	(10,230551.61)	(11,234751.62)	(12,237524.47)	(13,239361.87)	(14,240630.02)	(15,241495.8)	(16,242062.93)	(17,242451.56)	(18,242722.18)	(19,242913.04)	(20,243049.62)
			};
			\addlegendentry{\DA}
			\draw[densely dashed] ({rel axis cs:0,.1} -| {axis cs:6,0}) -- ({rel axis cs:0,1} -| {axis cs:6,0});
			\draw[densely dashed] ({rel axis cs:0,.3} -| {axis cs:7,0}) -- ({rel axis cs:0,1} -| {axis cs:7,0});
			\draw[densely dashed] ({rel axis cs:0,.5} -| {axis cs:9,0}) -- ({rel axis cs:0,1} -| {axis cs:9,0});
			\node at ({rel axis cs:0,.1} -| {axis cs:6,0}) [anchor=west] {$K(\ABM,\DA) = 6$};
			\node at ({rel axis cs:0,.3} -| {axis cs:7,0}) [anchor=west] {$K(\BM,\DA) = 7$};
			\node at ({rel axis cs:0,.5} -| {axis cs:9,0}) [anchor=west] {$K(\BM,\ABM) = 9$};
\end{axis}
\end{tikzpicture}
}

\newcommand{\PlotsRankDistStylized}{
\begin{tikzpicture}[scale=1]
\begin{axis}[
			mark size=1.5,
			xlabel={Rank}, 
			ylabel={Number of students}, 
			width=9cm,
			height=6cm,
			xtick=data,
			yticklabels={},
			legend style={at={(.02,.98)}, anchor=north west},
			]
			\addplot[color=BrickRed,mark=x] coordinates { 
				(1,15)	(2,30)	(3,35)	(4,40)	(5,45)	(6,50)	(7,54)
			};
			\addlegendentry{\BM}
			\addplot[color=blue,mark=square] coordinates { 
				(1,15)	(2,23)	(3,31)	(4,38)	(5,45)	(6,53)	(7,57)
			};
			\addlegendentry{\ABM}
			\addplot[color=OliveGreen,mark=o] coordinates { 
				(1,5)	(2,19)	(3,31)	(4,40)	(5,50)	(6,56)	(7,60)
			};
			\addlegendentry{\DA}
			\draw[densely dashed] ({rel axis cs:0,.2} -| {axis cs:3,0}) -- ({rel axis cs:0,1} -| {axis cs:3,0});
			\draw[densely dashed] ({rel axis cs:0,.4} -| {axis cs:4,0}) -- ({rel axis cs:0,1} -| {axis cs:4,0});
			\draw[densely dashed] ({rel axis cs:0,.6} -| {axis cs:5,0}) -- ({rel axis cs:0,1} -| {axis cs:5,0});
			\node at ({rel axis cs:0,.1} -| {axis cs:2.3,0}) [anchor=west] {$K(\ABM,\DA) = 3$};
			\node at ({rel axis cs:0,.3} -| {axis cs:3.3,0}) [anchor=west] {$K(\BM,\DA) = 4$};
			\node at ({rel axis cs:0,.5} -| {axis cs:4.3,0}) [anchor=west] {$K(\BM,\ABM) = 5$};
\end{axis}
\end{tikzpicture}
}

\onehalfspace
\def\bf{\normalfont\bfseries}
\changenotsign

\begin{document}

{\setstretch{1}
\title{
\LARGE{%
Trade-offs in School Choice: Comparing Deferred Acceptance, the Classic and the Adaptive Boston Mechanism}\thanks{\scriptsize{
Department of Informatics, University of Zurich, 8050 Zurich, Switzerland, \{mennle, seuken\}@ifi.uzh.ch. 
For updates see www.ifi.uzh.ch/ce/publications/ABM.pdf.
We would like to thank (in alphabetical order)
Nick Arnosti, 
Christian Basteck, 
Umut Dur, 
Aris Filos-Ratsikas, 
Katharina Huesmann, 
Fuhito Kojima, 
Antonio Miralles, 
Dmitry Moor, 
Suan Sebasti\'an Pereyra
and 
Utku \"Unver. 
Furthermore, we are thankful for the feedback we received at MIP'14 in Berlin and numerous anonymous reviewers at EC'15 and EC'16. 
Any errors remain our own. 
This research was supported by the Hasler Foundation under grant \#12078 and the SNSF (Swiss National Science Foundation) under grant \#156836.
}}}
\author{Timo Mennle \\ University of Zurich \and Sven Seuken\\ University of Zurich}
\date{%
First version: February 17, 2014\\
This version: \today}
\maketitle

\vspace{-2em}

\begin{abstract}
The three most common school choice mechanisms are 
the \emph{Deferred Acceptance mechanism (DA)}, 
the classic \emph{Boston mechanism (BM)}, 
and a variant of the Boston mechanism where students automatically skip exhausted schools, 
which we call the \emph{adaptive Boston mechanism (ABM)}. 
Assuming truthful reporting, we compare student welfare under these mechanisms both from a conceptual and from a quantitative perspective: 
We first show that, BM \emph{rank dominates} DA whenever they are comparable; 
and via limit arguments and simulations we show that ABM yields intermediate student welfare between BM and DA. 
Second, we perform computational experiments with preference data from the high school match in Mexico City. 
We find that student welfare (in terms of \emph{rank transitions}) is highest under BM, intermediate under ABM, and lowest under DA. 
BM, ABM, and DA can thus be understood to form a hierarchy in terms of student welfare. 
In contrast, in \citep{MennleSeuken2017PSP_WP}, we have found that the same mechanisms also form a hierarchy in terms of incentives for truthtelling that points in the opposite direction. 
A decision between them therefore involves an implicit trade-off between incentives and student welfare.
\end{abstract}

}
\section{Introduction}
\label{SEC:INTRODUCTION}
Each year, millions of children enter a new public school. 
To accommodate the students' (or their parents') preferences over schools, administrators must devise \emph{school choice mechanisms}. 
These are procedures that determine an assignment of students to schools on the basis of the reported preferences. 
Since the seminal paper by \citet{AbdulSonmez2003SchoolChoiceAMechanismDesignApproach}, school choice mechanisms have attracted the attention of economists, and a growing body of research has had substantial impact on policy decisions. 
\subsection{Boston `versus' Deferred Acceptance Mechanism}
\label{SEC:INTRODUCTION:BMVSDA}
Two particular mechanisms have received the lion's share of the attention: 
the \emph{Boston mechanism} and the \emph{Deferred Acceptance mechanism}. 
Both mechanisms collect preference reports from the students and then assign seats to students in rounds. 

Under the classic \emph{Boston mechanism (BM)}, students first apply to their favorite school. 
If a school has sufficient capacity to accommodate all applications in the first round, all applications are accepted. 
Otherwise, the school accepts applications following some priority order until its capacity is exhausted, and then it rejects all remaining applications. 
Students who were rejected in the first round apply to their second-choice school in the second round. 
The process repeats until all students have received a school or all schools have reached capacity. 
The Boston mechanism is ubiquitous in school choice, e.g., in Spain \citep{Calsamiglia2014IllusionOfChoiceBCN}, Germany \citep{BastekHuesmannNax2015MIPSecondarySchoolsGermany}, and the United States \citep{ErginSonmez2006GamesUnderBostonSchoolChoice}. 

The main motivation for letting students choose the schools through a school choice mechanism is \emph{student welfare}. 
Popular measures for student welfare are the number of students who received their top choice or one of their top-$3$ choices. 
Intuitively, BM fares well on these measures provided that students submit their preferences truthfully. 
The mechanism assigns as many applicants as possible to their first choices, then does the same with second choices in the second round, and so on. 
The mechanism owes much of its popularity to the intuitive way in which it attempts to maximize these measures of student welfare. 
However, BM is susceptible to strategic manipulation. 
In particular, it was found to disadvantage honest participants and to have ambiguous welfare properties in equilibrium \citep{ErginSonmez2006GamesUnderBostonSchoolChoice,Abdulkadiroglu2015ExpandingChoiceInSchoolChoice}. 

Concerns about the manipulability of the Boston mechanism have led to its abandonment in some cities in the US and around the world. 
In England, ``first-preference-first'' mechanisms (essentially the Boston mechanism) were even declared illegal in 2007 because they were believed to give unfair advantage to more sophisticated students \citep{Pathak2013AERComparingMechanismsByVulnerability}. 
To address problems associated with manipulability, the \emph{Deferred Acceptance mechanism (DA)} has been proposed as an alternative. 
Under DA, students also apply to schools in rounds. 
However, the acceptance at any school is \emph{tentative} rather than final. 
If in any subsequent round a student applies to a school with no free capacity, she is not automatically rejected. 
Instead, she will be accepted at that school if another student who has been previously tentatively accepted at the same school has lower priority. 
In this case, the tentative acceptance of a student with lowest priority is revoked and this student enters the next round of the application process. 

On the one hand, DA makes truthful reporting a dominant strategy for students. 
On the other hand, given the true preferences, BM appears to produce assignments with better student welfare. 
In this paper, we provide a formal justification for the latter statement, which has remained elusive so far.
\subsection{The Adaptive Boston Mechanism}
\label{SEC:INTRODUCTION:ABM}
So far, research on the Boston mechanism has largely focused on the classic BM described above. 
However, the Boston mechanism is sometimes used in a subtly different fashion: 
instead of applying to their $k^{\text{th}}$ choice in the $k^{\text{th}}$ round, in each round students apply to their \emph{most-preferred school that still has available capacity}. 

For example, in the city of Freiburg, Germany, approximately \numprint{1000} students transition from primary schools to one of ten secondary schools each year. 
Initially, they are asked to apply to their first-choice school. 
If this application is successful, their assignment is finalized. 
Students whose applications were rejected, receive a list of schools that still have seats available and are asked to apply to one of these schools in the second round. 
This process repeats in subsequent rounds. 

The procedure in Freiburg resembles the Boston mechanism, except that it prevents students from applying to schools that have no more open seats. 
This alteration leads to the \emph{adaptive Boston mechanism} (\emph{ABM}). 
ABM eliminates the risk of ``wasting one round'' by applying to an already exhausted school. 
Most school districts in the German state of Nordrhein-Westfalen also use ABM, 
and it was used for admission to secondary schools in Amsterdam until 2014 (where it was replaced by DA in 2015 \citep{HaanGautierOosterbeekKlaauw2015PerformanceSchoolAssignmentMechanismsPractice}). 

On the one hand, ABM removes some obvious opportunities for manipulation that exist under BM.
On the other hand, a student can obtain her third choice in the second round, which may prevent another student from getting her second choice in that round.
Consequently, one would expect ABM to take an intermediate position \emph{between} DA and BM in terms of student welfare (given the true preferences). 
In this paper, we provide a formal justification for this intuition. 
\subsection{Random Priorities in School Choice}
\label{SEC:INTRODUCTION:COARSE_PRIOS}
In prior work about school choice mechanisms, most results have been obtained under the assumption that priorities are fixed and strict. 
However, as \citet{KojimaUnver2014BostonSchoolChoice} have pointed out, this assumption is almost always violated: 
Priorities are typically \emph{coarse} in practice, e.g., if they are only based on neighborhoods or siblings.
Recently, the role of priorities has been further de-emphasized; 
for example, walk-zone priorities in Boston were abandoned in 2013 \citep{DurDukePathakSoenmez2014DemiseOfWalkZone}. 
Coarse priorities put many students in the same priority class.
Ties must then be broken, which introduces randomness into the mechanism. 
In this paper, we explicitly model the randomness that arises from coarse priorities and random tie-breaking by considering the \emph{random} assignments that arise before the tie-breaker has been implemented. 

In some school choice markets there are no initial priorities at all but the priorities are randomly generated by a single uniform lottery. 
For example, in 1999, the 15 neighborhoods of the Beijing Eastern City District used BM to assign students to middle schools and priorities were determined by a single uniform lottery \citep{Lai2009AdverseEffectsOfErrorsInSchoolChoiceBejing}. 
Similarly, the second phase of the school choice procedure in New York City used Deferred Acceptance with priorities derived in this way \citep{PathakSethuraman2011LotteriesStudentAssignmentEquivalence}; and most cities in Estonia employ \DA\ with single uniform priorities for elementary school assignment \citep{Lauri2014MIPEstoniaElementarySchools}. 
While in most school choice markets, priorities are not solely based on a single uniform lottery, the coarse nature of priorities is arguably closer to this assumption than to the assumption of strict and fixed priorities. 
All of our findings in this paper hold \emph{at least} for single uniform priorities, but most of them generalize to arbitrary priority structures with or without randomization. 
\subsection{A Motivating Example}
\label{SEC:INTRODUCTION:MOTIVATING_EXAMPLE}
In this paper, we uncover the relationships of BM, ABM, and DA in terms of student welfare when students report their preferences truthfully, independent of incentives. 
To obtain an intuition about these relationships, consider a market with four students, conveniently named 1, 2, 3, 4, and four schools, named $a$, $b$, $c$, $d$, with a single seat each. 
Suppose that the students' preferences are 
\begin{equation}
	P_1 ~:~ a \succ b \succ c \succ d, 
	\hspace{2em}
	P_2,P_3 ~:~ a \succ c \succ b \succ d, 
	\hspace{2em}
	P_4 ~:~ b \succ a \succ c \succ d,
\end{equation}
where $P_i: x \succ y$ indicates that student $i$ prefers school $x$ to school $y$. 
Furthermore, suppose that priorities are determined by a single uniform lottery. 
If the students report truthfully, the probabilities of each student obtaining each of the seats are as follows: 
{\small \begin{center}
\begin{tabular}{|c||c|c|c|c||c|c|c|c||c|c|c|c|}
	\hline
	 & \multicolumn{4}{c||}{\BM} & \multicolumn{4}{c||}{\ABM} & \multicolumn{4}{c|}{\DA} \\
	\hline
	Student & \hspace{0.55em}$a$\hspace{0.55em} & \hspace{0.55em}$b$\hspace{0.55em} & \hspace{0.55em}$c$\hspace{0.55em} & \hspace{0.55em}$d$\hspace{0.55em} & \hspace{0.55em}$a$\hspace{0.55em} & \hspace{0.55em}$b$\hspace{0.55em} & \hspace{0.55em}$c$\hspace{0.55em} & \hspace{0.55em}$d$\hspace{0.55em} & \hspace{0.55em}$a$\hspace{0.55em} & \hspace{0.55em}$b$\hspace{0.55em} & \hspace{0.55em}$c$\hspace{0.55em} & \hspace{0.55em}$d$\hspace{0.55em} \\
	\hline
	\hline
	1	& $1/3$ & $0$ & $0$ & $2/3$	& $1/3$ & $0$ & $1/3$ & $1/3$ & $1/3$ & $1/4$ & $1/6$ & $1/4$ \\
	\hline
	2 \& 3	& $1/3$ & $0$ & $1/2$ & $1/6$	& $1/3$ & $0$ & $1/3$ & $1/3$	& $1/3$ & $1/24$ & $3/8$ & $1/4$ \\
	\hline
	4	& $0$ & $1$ & $0$ & $0$	& $0$ & $1$ & $0$ & $0$	& $0$ & $2/3$ & $1/12$ & $1/4$   \\
	\hline
\end{tabular}
\end{center}}
To compare these assignments, we consider the \emph{cumulative rank distributions} \citep{Featherstone2011RankBasedRefinementWP} (i.e., the expected numbers of students who receive one of their top-$k$ choices for $k=1,\ldots,4$). 
These are:
{\small \begin{center}
\begin{tabular}{|l||c|c|c|c|}
	\hline
	Mechanism & $k=1$ & $k=2$ & $k=3$ & $k=4$ \\
	\hline
	\hline
	\BM	& $2$ & $3$ & $3$ & $4$ \\
	\hline
	\ABM	& $2$ & $2+2/3$ & $3$ & $4$ \\
	\hline
	\DA	& $1+2/3$ & $2+2/3$ & $3$ & $4$ \\
	\hline
\end{tabular}
\end{center}}
Observe that ABM and BM assign the same number of first choices but ABM assigns strictly fewer top-2 choices. 
Thus, BM yields higher student welfare in the sense that its rank distribution first order-stochastically dominates the rank distribution of ABM. 
Similarly, DA assigns less first choices than ABM, but for ranks 2 through 4, their cumulative rank distributions coincide. 
Thus, the rank distribution of ABM first order-stochastically dominates the rank distribution of DA. 
Consequently, in this example BM dominates ABM which in turn dominates DA.
\subsection{Contributions}
\label{SEC:INTRODUCTION:CONTRIBUTIONS}
In this paper, we expose the hierarchical relationship between the three most common school choice mechanisms, BM, ABM, and DA, in terms of student welfare from a conceptual and from a quantitative perspective. 
Conceptually, we show that BM rank dominates DA whenever this comparison is possible (Theorem \ref{THM:BM_DOM_DA}), 
that ABM is not rank dominated by DA \emph{in the limit} (Theorem \ref{THM:ABM_DOM_DA_LIMIT}), 
and that BM rank dominates ABM much more frequently than vice versa (Section \ref{SEC:RANK_DOM:BM_ABM}). 
Quantitatively, we conduct computational experiments on preference data from the high school match in Mexico City (Section \ref{SEC:RANK_TRANS}), 
where we find that, given true preferences, BM assigns more top-$k$ choices than ABM or DA for $k\leq 5$ and that ABM assigns more top-$k$ choices than DA for $k \leq 7$. 

In \citep{MennleSeuken2017PSP_WP}, we showed that while ABM is not strategyproof, it has better incentive properties than BM. 
This formally captures the intuition that ``not letting students apply to exhausted schools'' improves incentives. 
Our findings complement this insight by a second hierarchy in terms of student welfare in the opposite direction. 
A decision between the three most common school choice mechanisms thus involves an implicit trade-off between incentives and student welfare: 
first, if strategyproofness is a strict requirement, DA is obviously the mechanism of choice. 
Second, when administratoris believe that not letting students apply to exhausted schools provides sufficient incentives to elicit truthful preferences, then ABM may be a design alternative because in this case it improves student welfare, relative to DA.
Finally, if the true preferences are known or if students can be expected to report their preferences truthfully (despite the absence of strategyproofness guarantees), then BM yields the highest student welfare. 
We do not advocate superiority of either mechanism; 
instead, our insights allow administrators to make a conscious and informed decision about the implicit trade-off that they face when choosing between BM, ABM, and DA.
\section{Related Work}
\label{SEC:RELATED}
The classical Boston mechanism has received significant attention because it is frequently used for the assignment of students to public schools in many school districts around the world.
The mechanism has been heavily criticized for its manipulability: 
For the case of strict priorities, \citet{AbdulSonmez2003SchoolChoiceAMechanismDesignApproach} showed that BM is neither strategyproof nor stable. 
They suggested the Deferred Acceptance mechanism \citep{GaleShapley1962CollegeAdmission} as an alternative that is stable and strategyproof for students.
\citet{ErginSonmez2006GamesUnderBostonSchoolChoice} showed that \BM\ can have undesirable outcomes in full information equilibrium.
Experimental studies, such as those conducted by \citet{Chen2006SchoolChoiceExperiment} and \citet{PaisPinter2008SCandInforamtionExp}, revealed that it is indeed manipulated more frequently by human subjects than strategyproof alternatives.

\citet{KojimaUnver2014BostonSchoolChoice} provided an axiomatic characterization of the Boston mechanism for the case of fixed, strict priorities.
However, they also pointed out that the assumption of fixed, strict priorities is usually violated in school choice problems.
Some recent work has considered coarse priorities and revealed a number of surprising insights: 
\citet{Abdulkadiroglu2015ExpandingChoiceInSchoolChoice} demonstrated that in a setting with random priorities and perfectly correlated student preferences, BM can lead to higher \emph{ex-ante} welfare than Deferred Acceptance in equilibrium.
Similarly, simulations conducted by \citet{Miralles2008CaseForBostonWP} illustrated that with single uniform tie-breaking and no priorities, equilibria of the Boston mechanism can yield higher ex-ante welfare.
It has remained an open research question if and how the Boston mechanism can be understood to yield higher student welfare for general priority structures. 
Our present paper addresses this question: 
We show that, given truthful preferences, BM \emph{rank dominates} the Deferred Acceptance mechanism whenever the two mechanisms are comparable, 
and, using actual student preferences from the Mexico City high school match, we show that BM assigns substantially more top-$k$ choices than DA for $k\leq 7$. 

While the majority of prior work was focused on the classical Boston mechanism, the idea of an adaptive adjustment has previously been discussed as well. 
\citet{Alcalde1996ImplemetingStableSolutionsToMarriageProblems} studied a ``now-or-never'' mechanism for two-sided marriage markets, where men propose to their most-preferred \emph{available} partner in each round. 
\citet{Miralles2008CaseForBostonWP} informally argued that an adaptive order of applications may improve the position of unsophisticated (i.e., truthful) students. 
For the case when priorities are strict and fixed, \citet{Dur2015ModifiedBoston} provided an axiomatic characterization of the adaptive Boston mechanism. 
Furthermore, he showed that BM is at least \emph{as manipulable as} ABM in the sense of \citep{Pathak2013AERComparingMechanismsByVulnerability}.
If some students are unacceptable to some schools, then he also presented an example showing that BM is in fact \emph{more manipulable than} ABM. 
\section{Preliminaries}
\label{SEC:PRELIM}
\paragraph{Preferences.} 
Let $N$ be a set of $n$ \emph{students} and let $M$ be a set of $m$ \emph{schools}. 
We use $i$ to refer to particular students and $j$ or $a,b,c,\ldots$ to refer to particular schools. 
Each school $j$ has a \emph{capacity} of $q_j \geq 1$ seats, and we assume that there are enough seats to accommodate all students (i.e., $n \leq q_1 + \ldots + q_m$).\footnote{Otherwise, we can add a dummy school with capacity $n$ which all students rank last.} 
Students have strict \emph{preferences} $P_i$ over schools, where $P_i: a \succ b$ means that student $i$ \emph{prefers} school $a$ over school $b$. 
The set of all possible preference orders is denoted by $\mathcal{P}$.
A \emph{preference profile} ${P} = (P_1,\ldots,P_n) \in \mathcal{P}^N$ is a collection of preferences of all students, and we denote by $P_{-i}$ the collection of preferences of all students except $i$, so that we can write ${P} = (P_i,P_{-i})$. 

\paragraph{Priorities.}
A \emph{priority order} $\pi_j$ is a strict ordering of the students, where $\pi_j : i \succ i'$ means that student $i$ has priority over student $i'$ at school $j$, and we denote by $\Pi$ the set of all possible priority orders. 
A \emph{priority profile} is a collection of priority orders $\pi = \left(\pi_j\right)_{j \in M} \in \Pi^M$. 
$\pi$ is a \emph{single} priority profile if the priority orders of all schools are the same (i.e., $\pi_j = \pi_{j'}$ for all $j,j' \in M$), otherwise $\pi$ is a \emph{multiple} priority profile. 
A \emph{priority distribution} is a probability distribution $\mathds{P}$ over priority profiles.
$\mathds{P}$ is a \emph{single} priority distribution if it only randomizes over single priority profiles, or equivalently, $\text{supp}(\PP)$ contains only single priority profiles. 
A priority distribution $\mathds{P}$ \emph{supports all single priority profiles} if any single priority profile is chosen with strictly positive probability (i.e., $\mathds{P}[\pi] > 0$ for all single priority profiles $\pi$).
The uniform distribution over all single priority profiles is called the \emph{single uniform} priority distribution and denoted by $\mathds{U}$.

\paragraph{Assignments.}
A \emph{random assignment} is represented by an $n\times m$-matrix $x = (x_{i,j})_{i \in N, j \in M}$, where the entry $x_{i,j}$ is the probability that student $i$ receives a seat at school $j$. 
$x$ is \emph{feasible} if 
$x_{i,j} \in [0,1]$ for all $i,j$ (i.e., each entry is a probability), 
$\sum_{j \in M} x_{i,j} = 1$ for all $i$ (i.e., each student receives a seat with probability 1), 
and $\sum_{i \in N} x_{i,j} \leq q_j$ for all $j$ (i.e., no school's seats are assigned beyond capacity). 
$x$ is a \emph{deterministic assignment} if $x_{i,j} \in \{0,1\}$ for all $i,j$.\footnote{By virtue of the Birkhoff-von Neumann Theorem and its extensions \citep{Budishetal2013DesignRandomAllocMechsTheoryAndApp}, it suffices to consider the matrix representation as they can be represented as lotteries over deterministic assignments.}
Let $X$ and $\Delta(X)$ be the sets of all feasible deterministic and random assignments respectively.

\paragraph{Mechanisms.} 
A \emph{school choice mechanism} is a mapping $\varphi: \Pi^M \times \mathcal{P}^N \rightarrow X$ 
that receives as input a priority profile $\pp \in \Pi^M$ and a preference profile $ P \in \mathcal{P}^N$ and selects a deterministic assignment based on this input. 
For a given priority distribution $\mathds{P}$, the respective \emph{random school choice mechanism} (or just \emph{mechanism} for short) is the mapping $ \varphi^{\PP}: \mathcal{P}^N \rightarrow \Delta(X)$ that receives the students' preferences $ P$ and selects the random assignment by drawing a priority profile $\pi$ according to $\PP$ and applying $\varphi$ to $(\pi, P)$; 
formally, 
\begin{equation}
	\varphi^{\mathds{P}}( P) = \sum_{\pi \in \Pi^M} \varphi(\pi, P) \cdot \mathds{P}[\pi].
\label{EQ:CONSTUCTION_MECH_WITH_PRIO_DIST}
\end{equation}
If $\mathds{P}$ is a priority distribution that selects a particular priority profile $ \pi$ with certainty, then we may write $\varphi^{\pi}$ to denote the deterministic mechanism $\varphi^{\mathds{P}} = \varphi(\pi,\cdot)$. 
\paragraph{Common School Choice Mechanisms.}

In this paper, we consider the three most common school choice mechanisms, which assign students in rounds. 
Consider a given priority profile $\pi \in \Pi^M$ and a given preference profile $ P \in \mathcal{P}^N$. 

\emph{Deferred Acceptance mechanism (\DA).} 
In the first round of \DA\ each student $i$ applies to the school that is ranked first under $P_i$. 
Each school $j$ tentatively accepts applications according to $\pi_j$ until all seats at $j$ are filled, then $j$ rejects all remaining applications. 
In subsequent rounds all students who are not tentatively accepted at some school apply to their most-preferred school that has not rejected them yet. 
Then each school $j$ considers the set of students who are currently tentatively accepted at $j$ and the set of students who have newly applied to $j$ in this round. 
Of these students $j$ tentatively accepts applications according to $\pi_j$ until all seats at $j$ are filled, then $j$ rejects all remaining applications. 
When no school receives a new application all tentative assignments are finalized. 
Note that a student who has been tentatively accepted at some school, may be rejected from this school in a later round if another student with higher priority applies at that school. 

\emph{Classical Boston mechanism (\BM).} 
The first round of \BM\ is the same as under DA, except that the acceptances are final (not tentative). 
In round $k$, all students who have not been accepted at any school in previous rounds, apply to their $k^{\text{th}}$-choice school, and schools accept students into remaining open seats according to priority. 
Since acceptances are final, no student who has been accepted at a school, will be displaced by other students in later rounds, even if they have priority. 

\emph{Adaptive Boston mechanism (\ABM).} 
\ABM\ works like \BM, except that in each round students apply to their most-preferred school that still has at least one open seat. 
For example, if a student's second-choice school has filled up in the first round, this student applies to her third choice in the second round instead. 
Given any priority distribution $\mathds{P}$, we construct the corresponding random mechanisms $\DAP$, $\BMP$, and $\ABMP$ according to (\ref{EQ:CONSTUCTION_MECH_WITH_PRIO_DIST}). 
%
%
%
%
%
%
%
%
%
%
%
%
%
%
%
%
%
%
%
%
%
%
%
%
%
%
%
%
%
%
%
%
%
%
%
%
%
%
%
%
%
%
%
%
%
%
%
%
%
%
%
%
%
%
%
%
%
%
%
%
%
%
%
%
%
%
%
%
%
%
%
%
%
%
%
%
%
%
%
%
%
%
%
%
%
%
%
%
%
%
%
%
%
%
%
%
%
%
%
%
%
%
%
%
\section{Student Welfare Comparison by Rank Dominance}
\label{SEC:RANK_DOM}
Student welfare in school choice mechanisms is usually assessed using \emph{efficiency notions}. 
However, these notions do not allow for a meaningful comparison of \BM, \ABM, and \DA\ (Section \ref{SEC:RANK_DOM:CHALLENGES}). 
Instead, we use \emph{rank dominance} to establish the hierarchy in terms of student welfare (Sections \ref{SEC:RANK_DOM:BM_DA}, \ref{SEC:RANK_DOM:ABM_DA}, and \ref{SEC:RANK_DOM:BM_ABM}). 
\subsection{Challenges in Comparing Student Welfare}
\label{SEC:RANK_DOM:CHALLENGES}
Table \ref{TBL:EFF_COMPARISON} provides an overview of how \BM, \ABM, and \DA\ compare by different efficiency notions. 
We observe that all three mechanisms largely look alike: 
They produce ex-post efficient assignments (except $\DAP$ when multiple priority profiles are possible)
but neither of them is also ordinally efficient. 
Moreover, as we show in Appendix \ref{APP:EFF_FAILURE}, for the single uniform priority distribution $\mathds{U}$, none of the mechanisms lie on the efficient frontier.
The fact that traditional efficiency notions fail to differentiate between \BM, \ABM, and \DA\ necessitates our comparison in terms of rank distributions. 
\begin{table}%
\begin{tabular}{|l|c|c|c|c|}
\hline
Mechanism			& Ex-post efficient		& Ordinally	efficient$^{**}$		& On efficient frontier$^{***}$	\\ 
\hline \hline
$\BMP$ & \cmark & \xmark & \xmark \\
$\ABMP$ & \cmark & \xmark & \xmark \\
$\DAP$ & \xmark/\cmark$^*$& \xmark & \xmark \\	
\hline
\end{tabular}
\caption{Comparison of \BM, \ABM, and \DA\ in terms of efficiency notions. 
$^*$: $\DAP$ is ex-post efficient for single priority distributions but not ex-post efficient in general. 
$^{**}$: For non-deterministic priorities (see Appendix \ref{APP:ORDINAL_FAILURE}); for deterministic priorities, ex-post and ordinal efficiency coincide. 
$^{***}$: For the single uniform priority distribution $\mathds{U}$, neither mechanism is on the efficient frontier, subject to their respective incentive properties and symmetry (see Appendix \ref{APP:EFF_FAILURE}).}
\label{TBL:EFF_COMPARISON}
\end{table}%
\subsection{Rank Distributions and Rank Dominance}
\label{SEC:RANK_DOM:DEF}
The \emph{rank distribution} of an assignment consists of the numbers of students who receive their first, second, third, etc. choices. 
It is frequently used as a welfare criterion (e.g., see \citep{AbdulPathakRoth2009SPvsEffWithIndifferencesNYCSchoolMatch,Budish2012TheMultiUniAssignmentProblemTheoryAndEvidence}). 
\begin{definition}[Rank Distribution and Rank Dominance] 
\label{DEF:RANK_DIST_DOM}
Consider a preference profile $ P$ and two assignments $x,y \in \Delta(X)$. 
Let $\text{rank}_{ P}(i,j)$ denote the \emph{rank} of school $j$ in $P_i$ (i.e., the number of schools that student $i$ weakly prefers to $j$). 
Let 
\begin{equation}
	d_l^x = \sum_{i \in N, ~ j\in M:~rank_{ P}(i,j) = l} x_{i,j}%
\end{equation}%
be the expected number of students who receive their $l^{\text{th}}$ choice under $x$ and let $c_k^x = \sum_{l = 1}^k d_l^x$ be the expected number of students who receive one of their top-$k$ choices under $x$. 
The vector $c^{x} = (c_1^x,\ldots, c_m^x)$ is called the \emph{cumulative rank distribution of $x$ at $ P$}. 
We say that \emph{$x$ rank dominates $y$ at $ P$} if $c_k^x \geq c_k^y$ for all ranks $k \in \{1,\ldots,m\}$. 
This dominance is \emph{strict} if the inequality is strict for at least one rank $k$. 
\end{definition}
\citet{Featherstone2011RankBasedRefinementWP} formalized rank dominance as a way of comparing social welfare under ordinal assignment mechanisms. 
It captures the intuition that assigning \emph{two first and one second choices} is socially preferable to assigning \emph{one first and two second choices}. 
But of course, \emph{some} students may prefer their assignment under $y$ to their assignment under $x$, even if $x$ rank dominates $y$. 
Thus, choosing a rank dominant $x$ is not necessarily a Pareto improvement. 
Instead, it can be thought of as a \emph{tough decision} taken by the administrator: 
By choosing a rank dominant assignment the overall rank distribution improves but some students receive assignments they like less.

In the following we compare the assignments $\BMP( P)$, $\ABMP( P)$, and $\DAP( P)$ by their rank dominance relation at $ P$, where $ P$ is the true preference profile. %
\subsection{Comparing \BM\ and \DA}
\label{SEC:RANK_DOM:BM_DA}
Our second main result in this paper is that \BM\ rank dominates \DA\ whenever the outcomes of the two mechanisms are comparable. 
\begin{theorem}
\label{THM:BM_DOM_DA}
\ThmNbmDomDAStatement
\end{theorem}
\begin{proof}[Proof Outline (formal proof in Appendix \ref{APP:PROOF:BM_DOM_DA})] 
For any fixed single priority profile $ \pi$, we first prove the following statement: 
If $\DApi$ rank dominates $\BMpi$ at some preference profile, the assignments from both mechanisms must coincide. 
This insight allows us to formulate an \emph{extension} argument and obtain rank dominance for any single priority \emph{distribution}. 
For the case of fixed \emph{multiple} priority profiles, \citet{Harless2015ImmediateAcceptancePlus} proved the statement about coincidence of the assignments. 
Observing that his proof is essentially analogous to ours, we can re-use our extension argument to obtain rank dominance of $\BMP$ over $\DAP$ for \emph{arbitrary priority distributions} $\mathds{P}$. 
\end{proof}
Theorem \ref{THM:BM_DOM_DA} establishes a formal understanding of our intuition that, given truthful preferences, \BM\ produces the more appealing assignments: 
If the assignments from \BM\ and \DA\ are comparable in terms of rank dominance, then the assignment from \BM\ yields higher student welfare. 
The significance of Theorem \ref{THM:BM_DOM_DA} is further emphasized by the fact that traditional notions of efficiency are unable to differentiate between the two mechanisms, as shown in Section \ref{SEC:RANK_DOM:CHALLENGES}. 
\paragraph{Simulations.}
It is not difficult to construct an example where \BM\ and \DA\ are not comparable by rank dominance (see Example \ref{EX:BM_DA_INCOMPARABLE} in Appendix \ref{APP:EXAMPLE:BM_DA_INCOMPARABLE}). 
Therefore, the impact of Theorem \ref{THM:BM_DOM_DA} for practical market design depends on the frequency with which the assignments from \BM\ and \DA\ are comparable. 
We conduct simulations to estimate this frequency. 
Specifically, we consider settings with $m=10$ schools, $n \in \{ 50,100, \ldots,950,\numprint{1000}\}$ students, and evenly distributed capacities $q_j = q = n/m, j \in M$. 
In each setting we sample \numprint{10000} preference profiles $ P$ uniformly at random and we sample a single priority profile $ \pi$ for each $ P$. 
We then compute the assignments $\BMpi( P)$ and $\DApi( P)$ and compare them by their rank distribution.\footnote{By fixing a priority profile $\pi$ the simulations show the rank dominance relation from an \emph{ex-post} perspective. 
We take this approach because it is computationally infeasible to compute the random assignments from $\BMU$, $\ABMU$, and $\DAU$ for settings of any meaningful size.} 
By Theorem \ref{THM:BM_DOM_DA} there are three possible cases: 
(1) $\BMpi( P)$ strictly rank dominates $\DApi( P)$, 
(2) $\BMpi( P)$ and $\DApi( P)$ have the same rank distribution, 
or (3) $\BMpi( P)$ and $\DApi( P)$ are incomparable by rank dominance. 

\begin{figure}%
\begin{center}
\PlotsNbmPiDaPiRankDomComparison
\end{center}
\caption{Comparison of $\BMpi( P)$ and $\DApi( P)$ by rank dominance for \numprint{10000} preference profiles $ P$ sampled uniformly at random in settings with $m=10$ schools, $n \in \{50,100,\ldots,\numprint{1000}\}$ students, $q_j=n/m$.}%
\label{FIG:BMPI_DAPI_RANK_DOM_COMP}%
\end{figure}
The results of our simulation are shown in Figure \ref{FIG:BMPI_DAPI_RANK_DOM_COMP}. 
The share of preference profiles where $\BMpi$ and $\DApi$ are comparable by rank dominance (cases (1) \& (2)) is 24\% for smaller settings and $8\%$ for the largest setting ($n=\numprint{1000}$). 
Moreover, case (2) (where the rank distributions are exactly the same) occurs with a frequency of less than $0.1\%$ in settings with 250 or more students. 
Thus, in a city like Freiburg, Germany, with 10 schools and \numprint{1000} students, we would expect \BM\ to produce an assignment with unambiguously and strictly higher student welfare than \DA\ in 8\% of the cases (assuming uniformly distributed and truthful preferences). 

\medskip
The advantage of \BM\ in terms of student welfare contrasts with the advantage of \DA\ in terms of incentives. 
In this sense, Theorem \ref{THM:BM_DOM_DA} and Figure \ref{FIG:BMPI_DAPI_RANK_DOM_COMP} identify a \emph{cost of strategyproofness} that one incurs when choosing Deferred Acceptance over the classical Boston mechanism because strategyproofness is a strict requirement. 
Put differently, our findings demonstrate the improvements that \BM\ would yield if students reported their preferences truthfully, despite the absence of strategyproofness guarantees. 
\subsection{Comparing \ABM\ and \DA}
\label{SEC:RANK_DOM:ABM_DA}
Towards comparing student welfare under \ABM\ and \DA\ one can again attempt a comparison by the rank dominance relation. 
The motivating example from the introduction shows that $\ABMU$ can strictly rank dominate $\DAU$ at some preference profile. 
However, surprisingly, the opposite may also hold, as the following example shows. 
\begin{example}
\label{EX:DA_DOM_ABM}
Consider a setting with six students $N=\{1,\ldots,6\}$, six schools $M=\{a,\ldots,f\}$ with a single seat each, and the preference profile $ P = (P_1, \ldots, P_6)$ with 
\begin{equation*}
	P_1, \dots, P_4~:~ a \succ b \succ c \succ d \succ e \succ f \ \ \ \ \text{ and }\ \ \ \ P_5,P_6~:~e \succ b \succ a \succ d \succ f \succ c. 
\end{equation*}
It is straightforward to compute the assignments $\ABMU( P)$ and $\DAU( P)$. 
%
%
%
%
%
%
%
The corresponding cumulative rank distributions are 
$c^{\ABMU( P)} = (2, 3, 4, 5, 5, 6)$ 
and $c^{\DAU( P)} = (2, 3, 4, 5, 5 + 1/3, 6)$. 
Thus, $\DAU( P)$ strictly rank dominates $\ABMU( P)$ at $ P$.
\end{example}
From Example \ref{EX:DA_DOM_ABM} it is clear that a statement analogous to Theorem \ref{THM:BM_DOM_DA} does not hold between \ABM\ and \DA.\footnote{We can replace $\UU$ in Example \ref{EX:DA_DOM_ABM} by the single priority profile $ \pi : 1 \succ \ldots \succ 6$ and observe that $\DApi( P)$ still strictly rank dominates $\ABMpi( P)$ at $ P$.  
Thus, the comparison between ABM and DA in terms of rank dominance remains ambiguous, even if we restrict attention to fixed single priority profiles.}
To recover a meaningful comparison, we resort to two limit arguments. 
We consider two notions in which markets \emph{get large}: 
The first resembles large school choice markets, where the number of schools remains constant but the number of students and the schools' capacities grow. 
The second resembles large house allocation markets, where each school has a single seat, but the number of students and schools grow. 
For both notions we show that the share of preference profiles where $\DAU$ rank dominates $\ABMU$ (even weakly) vanishes in the limit. 
\begin{theorem}
\label{THM:ABM_DOM_DA_LIMIT}
\ThmDaNotDomAbmLimitStatement
\end{theorem}
\begin{proof}[Proof Outline (formal proof in Appendix \ref{APP:PROOF:ABM_DOM_DA_LIMIT})] 
For case \ref{THM:ABM_DOM_DA_LIMIT:SC}, we prove the stronger statement that the share of preference profiles where $\DAU$ assigns the same number of first choices as $\ABMU$ converges to zero. 
We give a bound for this share in terms of multinomial coefficients. 
Here, we must separately treat the conditional probabilities of the different cases that schools are un-demanded, under-demanded, over-demanded, or exactly exhausted as first choices. 
Case (\ref{THM:ABM_DOM_DA_LIMIT:HA}) follows in a similar fashion, but the proof is more involved as it requires the new notion of \emph{overlap} for preference profiles. 
We establish an upper bound for the share of profiles using 2-associated Stirling numbers of the second kind and variants of the Stirling approximation. 
\end{proof}
The surprising finding that $\DAU$ may strictly rank dominate $\ABMU$ (Example \ref{EX:DA_DOM_ABM}) challenges our initial hypothesis that $\ABMU$ yields higher student welfare.
However, Theorem \ref{THM:ABM_DOM_DA_LIMIT} alleviates this concern at least from a large market perspective.%
\footnote{Interestingly, complete enumeration reveals that $\DAU$ does not strictly rank dominate $\ABMU$ for \emph{any} setting with less than 6 schools (with unit capacity). 
In this sense, Example \ref{EX:DA_DOM_ABM} is minimal.} 

\paragraph{Simulations.}
The value of Theorem \ref{THM:ABM_DOM_DA_LIMIT} for practical market design again depends on the frequency of preference profiles where \ABM\ rank dominates \DA\ and the frequency of those preference profiles where \DA\ rank dominates \ABM\ (the surprising case from Example \ref{EX:DA_DOM_ABM}) in settings of meaningful size. 
To estimate these frequencies we conduct simulations analogous to those in Section \ref{SEC:RANK_DOM:BM_DA}. 
\begin{figure}%
\begin{center}
\PlotsAbmPiDaPiRankDomComparison
\end{center}
\caption{Comparison of $\ABMpi( P)$ and $\DApi( P)$ by rank dominance for \numprint{10000} preference profiles $ P$ sampled uniformly at random in settings with $m=10$ schools, $n \in \{50,100,\ldots,\numprint{1000}\}$ students, $q_j=n/m$.}%
\label{FIG:ABMPI_DAPI_RANK_DOM_COMP}%
\end{figure}
The results are shown in Figure \ref{FIG:ABMPI_DAPI_RANK_DOM_COMP}. 
First, we observe that $\DApi$ strictly rank dominates $\ABMpi$ for only 0.1\% of the preference profiles for $n=50$ students; 
for $n\geq 300$ students not a single such preference profile occurs in the sample. 
This suggests that the rate of convergence in Theorem \ref{THM:ABM_DOM_DA_LIMIT} is high and that the problematic cases where $\DApi$ rank dominates $\ABMpi$ are pathological and should not be a concern for practical market design. 
Second, we observe that $\ABMpi$ produces assignments with unambiguously higher student welfare for between 20\% of the preference profiles for smaller settings and 7\% for larger settings. 
For 10 schools and \numprint{1000} students, we would expect $\ABMpi$ to produce an assignment with unambiguously and strictly higher student welfare than \DA\ in 7\% of the cases (assuming uniformly distributed and truthful preferences). 

\medskip
Our limit result Theorem \ref{THM:ABM_DOM_DA_LIMIT} and our simulations identify an advantage of \ABM\ over \DA\ in terms of student welfare (given truthful preferences). 
Thus, a decision between these two mechanisms can be viewed as a trade-off between the full (instead of partial) strategyproofness of \DA\ and the higher student welfare under \ABM. 
For situations where partial strategyproofness \citep{MennleSeuken2017PSP_WP} is sufficient to elicit truthful preferences, these results demonstrate the improvements in student welfare that \ABM\ yields relative to \DA. 
\subsection{Comparing \BM\ and \ABM}
\label{SEC:RANK_DOM:BM_ABM}
Finally, we compare \BM\ and \ABM. 
The motivating example in the introduction illustrates that $\BMU$ may strictly rank dominate $\ABMU$. 
However, surprisingly, the opposite is also possible. 
\begin{example}
\label{EX:ABM_DOM_BM}
Consider a setting with five students $N=\{1,\ldots,5\}$, five schools $M=\{a,\ldots,e\}$ with a single seat each, and the preference profile $ P = (P_1, \ldots, P_5)$ with
\begin{equation*}
	P_1, P_2~:~ a \succ b \succ c \succ d \succ e,
	\ \ \ \ P_3,P_4~:~a \succ d \succ c \succ e \succ b,
	\ \ \ \ P_5~:~b \succ \ldots. 
\end{equation*}
%
%
%
%
%
%
%
%
%
As in Example \ref{EX:DA_DOM_ABM} we can compute the cumulative rank distributions $c^{\BMU( P)} = (2, 3, 4, 4+1/3, 5)$ and $c^{\ABMU( P)} = (2, 3, 4, 4+1/2, 5)$.
Since $4+1/2 > 4 + 1/3$, $\ABMU( P)$ strictly rank dominates $\BMU( P)$ at $ P$.\footnote{For the deterministic case the example holds with the single priority profile $\pi : 1 \succ 3 \succ 4 \succ 2 \succ 5$.}
\end{example}
Unfortunately, a theoretical result that establishes a rank dominance relationship between \BM\ and \ABM\ (even in the limit) appears currently out of reach. 
Therefore, we rely on simulations for this comparison. 
As in the comparison of \ABM\ and \DA\ in Section \ref{SEC:RANK_DOM:ABM_DA}, we want to estimate the frequencies of preference profiles where (1) $\ABMpi$ strictly rank dominates $\BMpi$ and those where (2) $\BMpi$ strictly rank dominates $\ABMpi$. 

\begin{figure}%
\begin{center}
\PlotsNbmPiAbmPiRankDomComparison
\end{center}
\caption{Comparison of $\BMpi( P)$ and $\ABMpi( P)$ by rank dominance for \numprint{10000} preference profiles $ P$ sampled uniformly at random in settings with $m=10$ schools, $n \in \{\numprint{50},\numprint{100},\ldots,\numprint{1000}\}$ students, $q_j=n/m$.}%
\label{FIG:BMPI_ABMPI_RANK_DOM_COMP}%
\end{figure}
The results are shown in Figure \ref{FIG:BMPI_ABMPI_RANK_DOM_COMP}. 
First, the share of the problematic cases (1) is below 0.6\% across all settings and reduces to 0.1\% for $n=\numprint{1000}$ students. 
Thus, albeit their occasional occurrence, cases like Example \ref{EX:ABM_DOM_BM} are the exception. 
Second, the share of preference profiles where $\BMpi$ strictly rank dominates $\ABMpi$ is between 21\% and 28\% across all settings.\footnote{For $n=\numprint{10000}$ students (not shown in Figure \ref{FIG:BMPI_ABMPI_RANK_DOM_COMP}) this share reduces to around 10\%. 
Based on this observation, we conjecture that it does converge to 0 as settings get larger, albeit slowly.} 
Thus, $\BMpi$ and $\ABMpi$ are frequently comparable and in these cases $\BMpi$ almost always rank dominates $\ABMpi$. 

These results show that we can expect \BM\ to produce assignments that are more appealing in terms of student welfare and that we can expect cases where the opposite is true to be very rare. 
Analogous to the \emph{cost of strategyproofness} we identified by comparing \BM\ and \DA\ (see Section \ref{SEC:RANK_DOM:BM_DA}), this difference in terms student welfare can be viewed as a \emph{cost of partial strategyproofness} when choosing \ABM\ over \BM. 
In other words, we identify the gains in student welfare that could be obtained by using \BM\ instead of \ABM\ if students reported their preferences truthfully, despite the absence of any strategyproofness guarantees. 

\medskip
In summary, in Section \ref{SEC:RANK_DOM}, we have shown that whenever \BM, \ABM, and \DA\ are comparable by rank dominance, then the rank dominance relation generally points in the expected direction with some notable exceptions. 
Thus, we have established a hierarchy between \BM, \ABM, and \DA\ in terms of student welfare (given truthful preferences), which is the exact inverse of the hierarchy in terms of incentives (from \citep{MennleSeuken2017PSP_WP}).
\section{Student Welfare Comparison by Rank Transitions}
\label{SEC:RANK_TRANS}
So far, we have compared \BM, \ABM, and \DA\ by rank dominance. 
However, we have also observed that they are not comparable in this way for many preference profiles. 
For this reason we now introduce \emph{rank transitions} and apply this concept to obtain a more general albeit slightly weaker comparison of \BM, \ABM, and \DA. 
\subsection{Rank Transitions}
\label{SEC:RANK_TRANS:RT}
First, we introduce \emph{rank transitions}, a generalization of rank dominance. 
The notion of rank transitions is motivated by a common welfare criterion: 
In many school choice markets it is the express objective to assign as many students as possible to their first choice, to one of their top-3 choices, or generally to one of the their top-$k$ choices. 
Rank transitions capture this criterion formally. 
\begin{definition}[$K$-Rank Dominance and Rank Transition]
\label{DEF:K_RANK_DOM}
Consider a preference profile $ P$ and two assignments $x,y\in \Delta(X)$ with cumulative rank distributions $c^x$ and $c^y$ at $ P$ respectively. 
For any rank $K \in \{1,\ldots,m\}$ we say that \emph{$x$ $K$-rank dominates $y$ at $ P$} if $c^x_k \geq c^y_k$ for all ranks $k \in \{1,\ldots,K\}$. 
The largest rank $K$ for which $x$ $K$-rank dominates $y$ at $ P$ is called the \emph{rank transition (between $x$ and $y$ at $ P$)}.
\end{definition}

\begin{figure}
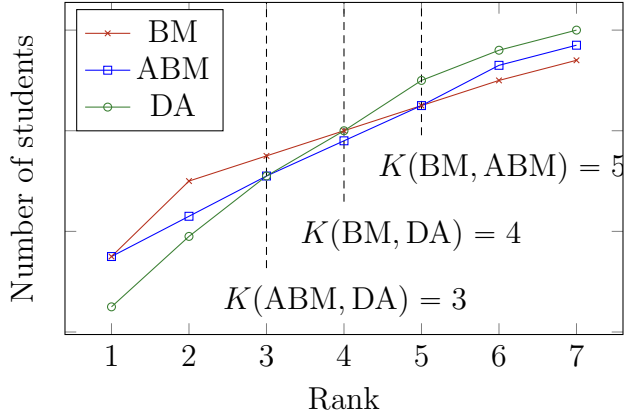
%
\begin{center}
\PlotsRankDistStylized
\end{center}
\caption{Stylized example of cumulative rank distributions and rank transitions. }%
\label{FIG:EX_RANK_DIST}%
\end{figure}
$K$-rank dominance of an assignment $x$ over another assignment $y$ means that $x$ rank dominates $y$ \emph{up to rank $K$}. 
For example, if our primary objective is to assign as many students as possible to their top-$3$ choices, then we prefer a $3$-rank dominant assignment. 
Conversely, if the rank transition between $x$ and $y$ is $5$, then we prefer $x$ over $y$ if we care specifically to assign as many students as possible to their top-$k$ choices for any $k\leq 5$. 
A higher rank transition between two mechanisms therefore implies a stronger differentiation between them in terms of student welfare, provided that we care about top-$k$ choices. 

Figure \ref{FIG:EX_RANK_DIST} shows stylized rank distributions for \BM, \ABM, and \DA. 
In this example, the rank transition between \ABM\ and \DA\ lies at $K(\ABM,\DA)=3$. 
This means that \ABM\ assigns (weakly) more top-1, top-2, and top-3 choices than \DA, but fails to assign more top-4 choices. 
The rank transitions between \BM\ and \DA\ and between \BM\ and \ABM\ lie at $K(\BM,\DA)=4$ and $K(\BM,\ABM)=5$ respectively. 
\subsection{Influence of Setting Parameters on Rank Transitions}
\label{SEC:RANK_TRANS:INFLUENCE}
\begin{figure}
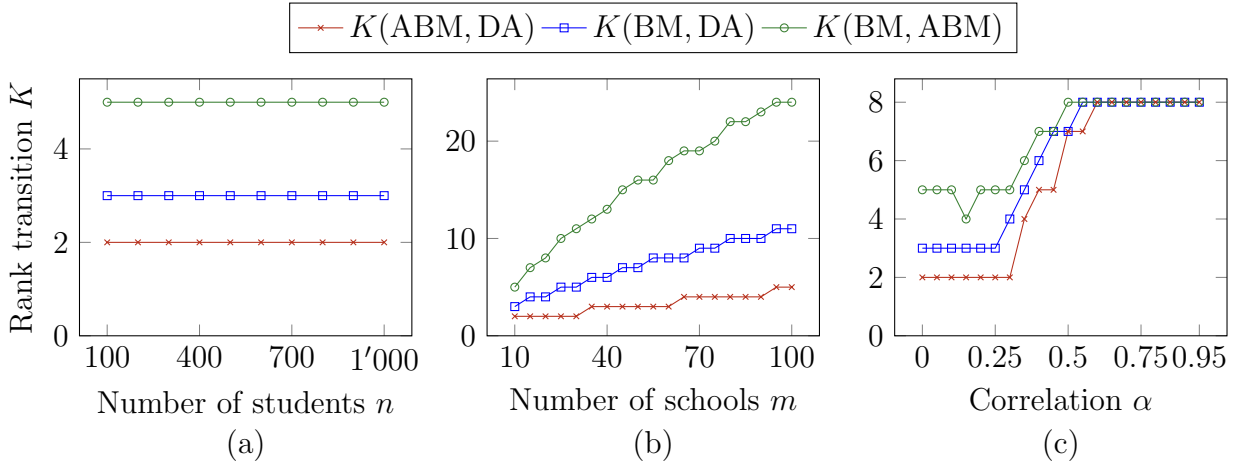
%
\begin{center}
\PlotsTopKFactorEffects
\end{center}
\caption{Influence of setting parameters $n$ (a), $m$ (b), $\alpha$ (c) on rank transitions}%
\label{FIG:TOPK_EFFECTS}%
\end{figure}

In this section, we study how variations in the setting parameters influence the rank transitions between \BM, \ABM, and \DA. 
Specifically, we consider 
the number of students $n$, 
the number of schools $m$, 
and correlation in the students' preferences $\alpha$.  
\citet{Arnosti2016CentralizedClearinghousesTradeoff} proved single crossing properties that hold in continuum markets.  
These results do not apply directly to the finite settings that we study. 
However, based on \citeauthor{Arnosti2016CentralizedClearinghousesTradeoff}'s insights for continuum markets, we formulate four hypotheses: 
\begin{description}
	\item[Hypothesis 1] 
		\label{HYP:RT_INDEPT_N} 
		Holding $m$ fixed and adjusting capacities to accommodate all students, rank transitions are independent of $n$.
	\item[Hypothesis 2] 
		\label{HYP:RT_DEPT_M} 
		Holding $n$ fixed and adjusting capacities to accommodate all students, rank transitions increase with $m$.
	\item[Hypothesis 3] 
		\label{HYP:RT_DEPT_ALPHA} 
		Holding $n$, $m$, and capacities fixed, rank transitions increase with $\alpha$. 
	\item[Hypothesis 4] \label{HYP:RT_SORTED} The order of the rank transitions is as shown in Figure \ref{FIG:EX_RANK_DIST}: 
		\begin{equation}
			K(\ABM,\DA) \leq K(\BM,\DA) \leq K(\BM,\ABM).
		\end{equation}
\end{description}
We assess these hypotheses via simulations. 
Figure \ref{FIG:TOPK_EFFECTS} shows plots of estimated rank transitions for varying setting parameters. 

Towards Hypothesis 1 we consider settings with $m=10$ schools, $n\in \{\numprint{100},\numprint{200},\ldots,\numprint{1000}\}$ students, and capacities $q_j = q =n/m$. 
In each setting we sample \numprint{100000} (uncorrelated) preference profiles $ P$ uniformly at random and a single priority profile $\pi$ for each $ P$. 
We estimate the expected rank distribution by averaging the rank distributions across all samples. 
Then we compute the rank transitions $K(\ABM,\DA)$, $K(\BM,\DA)$, and $K(\BM,\ABM)$. 
As we see in Figure \ref{FIG:TOPK_EFFECTS} (a), the rank transitions are constant at $K(\ABM,\DA)=2$, $K(\BM,\DA)=3$, and $K(\BM,\ABM)=5$ across all settings, which is in line with Hypothesis 1. 

Towards Hypothesis 2 we consider settings with $m\in \{\numprint{10},\numprint{15},\ldots,\numprint{100}\}$ schools, capacities $q_j = q =n/m$, and $n \approx \numprint{1000}$ students (if $n=\numprint{1000}$ is not divisible by $m$, we reduced $n$ to the highest multiple of $m$ below \numprint{1000}). 
Figure \ref{FIG:TOPK_EFFECTS} (b) shows the estimated rank transitions. 
Consistent with Hypothesis 2, they grow as the number of schools increases. 

Towards Hypothesis 3 we consider settings with $n = \numprint{1000}$ students, $m = \numprint{10}$ schools, and capacities $q_j = \numprint{100}$. 
We correlate the students' preferences by sampling vNM utility profiles with common and private value components, where $\alpha \in \{\numprint{0.00},\numprint{0.05},\ldots,\numprint{0.90},\numprint{0.95}\}$ is the share of the common value component in each student's utility function.\footnote{This approach follows prior work \citep{ErdilErgin2008WhatsTheMatterWithTieBreaking,%
Budish2012TheMultiUniAssignmentProblemTheoryAndEvidence,%
Abdulkadiroglu2015ExpandingChoiceInSchoolChoice,
MennleSeukenWeissPhilipp2015LocalManipulationStrategies}: 
We draw common values $(v_j)_{j\in M}$ from $U[0,1]^M$ and for each student $i$ we draw private values $(w_{i,j})_{j\in M}$ from $U[0,1]^M$. 
Student $i$'s utility for school $j$ is given by $u_i(j) = \alpha v_j + (1-\alpha) w_{i,j}$.} 
Figure \ref{FIG:TOPK_EFFECTS} (c) shows the estimated rank transitions. 
Again, we observe that they grow as the correlation $\alpha$ increases, which is consistent with Hypothesis 3 (with the exception of $K(\BM,\ABM)=4$ for $\alpha = 0.15$). 

Finally, observe that in all three plots, the order of the rank transitions is the same for all settings and consistent with Hypothesis 4, namely that the order of the rank transitions is $K(\ABM,\DA) \leq K(\BM,\DA) \leq K(\BM,\ABM)$. 

\medskip
These results show that rank transitions are a meaningful way to assess differences in student welfare under \BM, \ABM, and \DA: 
when the rank dominance comparison between them is inconclusive, we can still expect that \BM\ dominates \ABM\ which in turn dominates \DA\ \emph{up to their respective rank transitions}. 
However, the setting parameters $n$, $m$, and $\alpha$ are usually outside the control of the administrator. 
In order to understand the differences in student welfare that we expect in practice, we must measure the rank transitions on actual student preferences. 
\subsection{Rank Transitions in Practice: 
Evidence from the Mexico City High School Match}
\label{SEC:RANK_TRANS:MEX}
In this section, we present results from computational experiments with student preference data from high school admissions in Mexico City from years 2010 through 2014. 
We find that student welfare in terms of rank transitions is highest under \BM, intermediate under \ABM, and lowest under \DA\ (given truthful preferences).

In Mexico City around \numprint{280000} students are assigned to around \numprint{600} educational options at public high schools each year. 
The assignment is centrally organized by Comisi\'on Metropolitana de Instituciones P\'ublicas de Educaci\'on Media Superior (COMIPEMS). 
In the \emph{Main Phase} students submit rank-ordered lists of up to 20 options. 
\DA\ is used to compute the assignment, where a standardized exam determines the (single) priority structure. 
Around 85\% of the students are assigned in the Main Phase. 
In a Second Phase the remaining 15\% are assigned to options with excess capacity in a decentralized process. 
The data for our computational experiments consists of the rank-ordered lists submitted by the students during the Main Phase in the years 2010 through 2014. 
Following prior work by \cite{AbdulPathakRoth2009SPvsEffWithIndifferencesNYCSchoolMatch}, we use the number of students assigned to each option during the Main Phase as a proxy for capacity.\footnote{We give a detailed description of the data and our analysis in Appendix \ref{APP:DATA_DESCRIPTION}. 
We are grateful to Juan Sebasti\'an Pereyra, who ran the experiment software for us on a proprietary computer system of ECARES at the Universit\'e Libre de Bruxelles in Brussles, Belgium.} 

We are interested in identifying the rank distributions and rank transitions that arise from applying $\BMU$, $\ABMU$, and $\DAU$ in the Main Phase of the Mexico City high school match. 
To estimate these we sample \numprint{2000} single priority profiles uniformly at random and average the resulting rank distributions. 
\begin{figure}
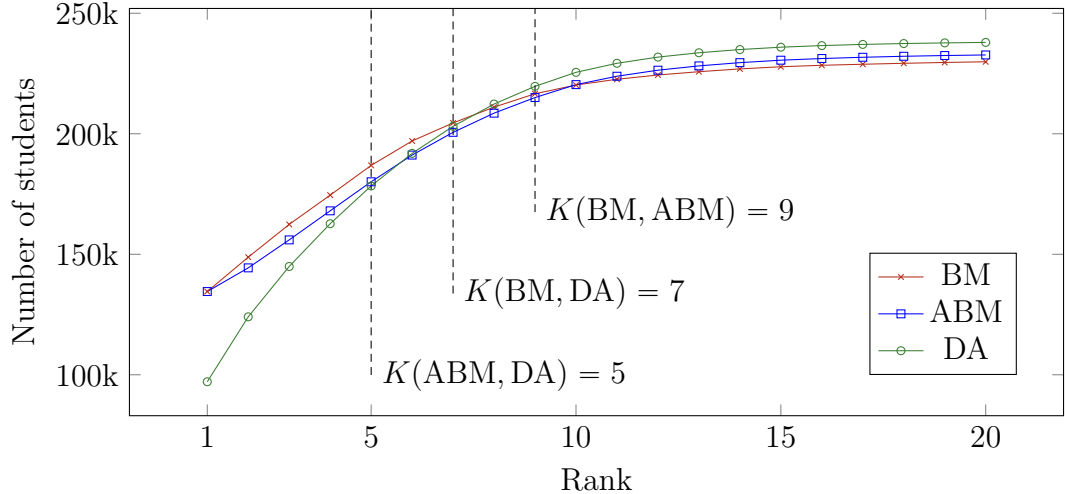
%
\begin{center}
\PlotsMexRankDistFourteenSingleMain
\end{center}
\caption{Rank distributions and rank transitions between $\BMU$, $\ABMU$, and $\DAU$ for student preferences from Mexico City high school match in 2014}%
\label{FIG:MEX_RANK_DIST}%
\end{figure}
Figure \ref{FIG:MEX_RANK_DIST} shows the results for the preference data from 2014. 
All estimated values are statistically different (paired t-test,\footnote{For each rank we use the Shapiro-Wilk test to ascertain normality of the sample differences. 
Subsequently we use the Student's t-test for paired samples to test for difference.} Student $p < 0.001$). 

First, we observe that the rank distributions have the single crossing structure that we hypothesized in Section \ref{SEC:RANK_TRANS:INFLUENCE}. 
Thus, rank transitions are indeed a meaningful measure of differences in student welfare. 
The rank transitions for 2014 are 
\begin{equation}
	K(\ABM,\DA) = 5\ <\ K(\BM,\DA) = 7\ <\ K(\BM,\ABM) = 9. 
\end{equation}
This means that while the assignments are not directly comparable by rank dominance, we see that $\BMU$ rank dominates $\DAU$ \emph{up to rank $K=7$}, $\ABMU$ rank dominates $\DAU$ \emph{up to rank $K=5$}, and $\ABMU$ is rank dominated by $\BMU$ \emph{up to rank $K=9$}. 
Thus, if we care about students' top-$k$ choices for $k \leq 5$, the three mechanisms form a clear hierarchy in terms of student welfare on the preferences of students from Mexico City, with \BM\ outperforming \ABM\ which in turn outperforms \DA. 

Second, we observe that on average $\DAU$ assigns \numprint{97103} students to their first choices (34.1\% of all students). 
In contrast, $\BMU$ and $\ABMU$ assign \numprint{134545} (47.3\%) first choices (i.e., \numprint{37442} more), a substantial improvement. 
At the same time, relative to $\DAU$, $\BMU$ and $\ABMU$ leave \numprint{8042} (2.8\%) and \numprint{5207} (1.8\%) additional students to be assigned in the Second Phase respectively. 
Thus, if we care most about first choices, $\BMU$ and $\ABMU$ vastly improve student welfare but at the cost of a small increase in the number of students who must be assigned in the Second Phase (and of course the fact that neither $\BMU$ nor $\ABMU$ are strategyproof). 
\begin{remark}
We have conducted the same analysis on the preference data from Mexico City for all years 2010--2014 and for the single and multiple uniform priority distribution. 
The findings are consistent across all years and priority distributions and are given in Appendix \ref{APP:DATA_DESCRIPTION}. 
\end{remark}
In summary, we find that when applied to student preferences from Mexico City, student welfare is highest under \BM, intermediate under \ABM, and lowest under \DA\ (given truthful preferences). 
This refines our understanding of the hierarchical relationship of the three mechanisms in terms of student welfare even if their assignments are not comparable by rank dominance. 
\section{Conclusion}
\label{SEC:CONCLUSION}
Among the three most common school choice mechanisms, DA, BM, and ABM, only DA is strategyproof for students, while ABM is at least partially strategyproof whereas BM is not \citep{MennleSeuken2017PSP_WP}. 
They therefore form a hierarchy in terms of their incentive properties. 
In the present paper, we have complemented this analysis by showing a second hierarchy in terms of student welfare points in the opposite direction. 

To this end, we have taken two complementary approaches: 
First, we have established the relationship between the three mechanisms in terms of rank dominance. 
When the comparison of the assignments from BM, ABM, and DA is possible, 
then it generally points in the expected direction (except in a very small number of cases). 
Using simulations, we have also established that the share of preference profiles where this comparison applies is sufficiently large to make the results relevant for practical market design. 
Second, we have compared the rank distributions on student preference data from the Mexico City high school match. 
Our computational experiments have shown that the hierarchy persists on that data. 
When we care about assigning top-$k$ choices for $k \leq 5$, BM is preferable to ABM which is preferable to DA (given truthful preferences). 

The general lesson to be learned from our results is that a decision between BM, ABM, and DA involves an implicit trade-off between incentives and student welfare. 
If strategyproofness is a strict requirement, then DA is obviously the mechanism of choice. 
When administrators are concerned about incentives but partial strategyproofness is sufficient to elicit truthful preferences, then ABM may be a design alternative because in this case it can improve student welfare. 
Finally, if the true preferences are known or if students can be expected to submit their preferences truthfully, despite the absence of strategyproofness guarantees, then BM yields the highest student welfare. 
Rather than advocate superiority of one of the mechanisms, we provide market designers with the means to make a conscious decision about this trade-off.

%


%
%
%
%
\appendix
\section*{APPENDIX}

\section{Proof of Theorem \ref{THM:BM_DOM_DA}}
\label{APP:PROOF:BM_DOM_DA}
\begin{proof}[Proof of Theorem \ref{THM:BM_DOM_DA}]
\ThmNbmDomDAProof
\end{proof}

\section{Example from Section \ref{SEC:RANK_DOM:BM_DA}}
\label{APP:EXAMPLE:BM_DA_INCOMPARABLE}
\begin{example}
\label{EX:BM_DA_INCOMPARABLE}
\ExNbmDaIncomparable
\end{example}

\section{Proof of Theorem \ref{THM:ABM_DOM_DA_LIMIT}}
\label{APP:PROOF:ABM_DOM_DA_LIMIT}
\begin{proof}[Proof of Theorem \ref{THM:ABM_DOM_DA_LIMIT}]
\ThmDaNotDomAbmLimitProof
\end{proof}

\section{Data from Mexico City High School Match, 2010--2014}
\label{APP:DATA_DESCRIPTION}
In this section we describe the matching procedure in Mexico City, the data we have used for our computational experiments, and the experiments themselves. 
In Mexico City around \numprint{280000} students are assigned to around \numprint{600} educational options at public high schools each year. 
The assignment is centrally organized by Comisi\'on Metropolitana de Instituciones P\'ublicas de Educaci\'on Media Superior (COMIPEMS). 

\paragraph{Matching Process (text adopted verbatim with only minor modifications from \citep{LiPereyra2015SelfSelection}).}
The matching proceeds as follows: 
\begin{itemize}
	\item In \emph{late January}, COMIPEMS hands out brochures disseminating information about available options and instructions about the matching process. 
		In addition, information about past assignments is available to students. 
	\item In \emph{March}, after registering with COMIPEMS, students submit their preferences over up to 20 options. 
		Each high school offers either a single option or multiple options. 
		Additionally, students have to fill out a survey questionnaire.
	\item In \emph{June}, all students simultaneously take a standardized exam. 
		The final score is used to determine student's priority in the match. 
		A minimum score of 31 out of 128 is required for eligibility. 
		This minimum score requirement was abandoned as of 2013 in an endeavor to extend compulsory education to
high school level. 
	\item In \emph{mid July}, students have to provide their secondary school certificate to be eligible for the match.
	\item In \emph{late July}, the match is run in two phases:
		\begin{enumerate}
			\item \emph{Main Phase:} Based on the priority defined by exam scores, COMIPEMS allocates students using the \emph{Serial Dictatorship (SD)} mechanism, which is equivalent to the Deferred Acceptance mechanism for single priority orders, such as those based on exam score. 
				When students tie for the last seats at an option, COMIPEMS consults the school either to take all or reject all tied students.
			\item \emph{Second Phase:} Those students who have remained unassigned because all their ranked options are full are allowed to register in options with excess capacity. 
				This process is decentralized.
		\end{enumerate}
\end{itemize}

\paragraph{Description of the Data.}
The data for our computational experiments consists of the preference lists submitted by the students and the number of students who were assigned to each of the educational options in the Main Phase. 
\begin{table}%
\begin{center}
{\scriptsize\begin{tabular}{|l|c|c|c|c|c|}
\hline
\textbf{Year}		& \textbf{2010} & \textbf{2011} & \textbf{2012} & \textbf{2013} & \textbf{2014}  \\ 
\hline \hline
\# of options & \numprint{536} & \numprint{569} & \numprint{595} & \numprint{611} & \numprint{628} \\
\hline
\hline
\# of students & \numprint{276581} & \numprint{270699} & \numprint{269775} & \numprint{280001} & \numprint{284412} \\
\hline
\# of seats (estimate) & \numprint{230074} & \numprint{230194} & \numprint{234340} & \numprint{246317} & \numprint{247977} \\
\hline
Excess \# of students & \numprint{46507} (16.8\%) & \numprint{40505} (15.0\%) & \numprint{35435} (13.1\%) & \numprint{33684} (12.0\%) & \numprint{36435} (12.8\%) \\
\hline
\hline
Average list length & 9.7 & 9.9 & 10.1 & 9.9 & 9.8 \\
\hline
\# lists with length 20 & \numprint{9266} (3.4\%) & \numprint{9583} (3.5\%) & \numprint{10598} (3.9\%) & \numprint{10917} (3.9\%) & \numprint{11557} (4.1\%) \\
\hline
\end{tabular}}
\end{center}
\label{TBL:DATA_STATS}
\caption{Summary statistics of data from Mexico City high school match in year 2010--2014.}
\end{table}%
Table \ref{TBL:DATA_STATS} gives summary statistics of this data. 

The Main Phase uses the Deferred Acceptance mechanism with a cap of 20 on preference lists. 
This cap and the flexibility in the capacity constraints make the mechanism manipulable in theory. 
Nonetheless, for the purposes of our analysis, we treat the submitted preferences as approximately truthful. 
Two insights justify this assumption:
first, any non-truthful best response under DA with caps involves ranking a maximum number of options. 
However, only between 3.4\% and 4.1\% of the students ranked the maximum number of 20 options and on average they ranked between 9.7 and 10.1 options. 
If students had best-responded to the cap on preference lists, we would expect these values to be much higher. 
Second, to find a beneficial manipulation that exploits the flexibility of capacities, students need to know the priorities and preference reports of all other students. 
As students do not possess this information, such manipulations are unlikely to occur. 

For our analysis we consider only those students who submitted their secondary school certificate in mid July. 
This restriction is motivated by the fact that students who did not submit this certificate are not eligible for any public high school, even in the Second Phase. 
We do include students who submitted their certificate but whose exam score was too low for admittance to any of the options they ranked. 
Our motivation for including these students is that our goal is to compare the performance of $\BMU$, $\ABMU$, and $\DAU$ on actual student preferences in (hypothetical) settings with random priorities. 
Therefore, we disregard the fact that these students did not have sufficient priority in the particular situation of Mexico City. 

The exact number of seats in each option are unknown. 
Moreover, the way in which ties over last seats are broken induces some flexibility in the number of seats. 
As a proxy for capacities we therefore use the number of students that were assigned to each program in the Main Phase. 
As students are assigned to options with excess capacity in the Second Phase, this represents a pessimistic estimate. 
In particular the total number of students exceeds the estimated total number of seats by 12.0\% to 16.8\%. 
Thus, the excess students will remain unassigned in our computational experiments (and possibly more). 

\paragraph{Data Access.}
This data has previously been analyzed by \citet{LiPereyra2015SelfSelection}. 
It is protected under a non-disclosure agreement between COMIPEMS and the research team at ECARES. 
We are grateful to Juan Sebasti\'an Pereyra of the ECARES research team, who ran the experiment software for us on a proprietary computer system of ECARES at the Universit\'e Libre de Bruxelles in Brussles, Belgium. 
The experiment software compiled the summary results, which we required for our analysis. 
At no point did we, the authors of this paper, have access to any personally identifiable student information. 

\paragraph{Computational Experiments.}
For each year (i.e., 2010 through 2014) and each priority distribution (i.e., single and multiple uniform priority distribution) we have estimated cumulative rank distributions based on \numprint{2000} samples of $\BMpi( P)$, $\ABMpi( P)$, and $\DApi( P)$. 
For each rank $k \in \{1,\ldots,20\}$ we have used the Shapiro-Wilk test to verify that the differences between the samples $c_k^{\BMP( P)}$, $c_k^{\ABMP( P)}$, and $c_k^{\DAP( P)}$ are normally distributed. 
Then we have used Student's t-test to test the estimates for difference. 
For all years, priority distributions, and ranks, the estimated $\widehat{c_k^{\BMP( P)}}$, $\widehat{c_k^{\ABMP( P)}}$, and $\widehat{c_k^{\DAP( P)}}$ are significantly different for $p<10^{-15}$. 
\section{Complete Results of Computational Experiments}
\label{APP:ADDITIONAL_RESULTS}
Figures \ref{FIG:MEX_RANK_DIST_TEN_SING}--\ref{FIG:MEX_RANK_DIST_FOURTEEEN_MULT} show the cumulative rank distributions and rank transitions for the years 2010--2014 and for the single and multiple uniform priority distributions, analogous to Figure \ref{FIG:MEX_RANK_DIST} (which only showed the results for the single uniform priority distribution and preference data from 2014). 
Table \ref{TBL:RANK_TRANS_AVG_RANK_SINGLE_ALL} summarizes the rank transitions that we identified. 
\begin{table}
\begin{center}
\begin{tabular}{|l|l||c|c|c|}
\hline
\textbf{Years} & \textbf{Priority dist.} & $K(\ABM,\DA)$ & $K(\BM,\DA)$ & $K(\BM,\ABM)$ 	\\ 
\hline
\hline
2010, 2011, 2012 & single uniform & 5 & 7 & 10 \\
\hline
2013, 2014 & single uniform & 5 & 7 & 9 \\
\hline
\hline
2010, 2011 & multiple uniform & 7 & 8 & 10 \\
\hline
2012 & multiple uniform & 7 & 7 & 10 \\
\hline
2013, 2014 & multiple uniform & 6 & 7 & 9 \\
\hline
\end{tabular}
\end{center}
\label{TBL:RANK_TRANS_AVG_RANK_SINGLE_ALL}
\caption{Rank transitions for student preferences from Mexico City, 2010--2014, single and multiple uniform priority distributions.}
\end{table}%
\begin{figure}[H]%
\PlotsMexRankDistTenSingle
\caption{Rank distributions for Mexico City student preferences, single uniform priority distribution, 2010}%
\label{FIG:MEX_RANK_DIST_TEN_SING}%
\end{figure}
\begin{figure}[H]%
\PlotsMexRankDistElevenSingle
\caption{Rank distributions for Mexico City student preferences, single uniform priority distribution, 2011}%
\label{FIG:MEX_RANK_DIST_ELEVEN_SING}%
\end{figure}
\begin{figure}[H]%
\PlotsMexRankDistTwelveSingle
\caption{Rank distributions for Mexico City student preferences, single uniform priority distribution, 2012}%
\label{FIG:MEX_RANK_DIST_TWELVE_SING}%
\end{figure}
\begin{figure}[H]%
\PlotsMexRankDistThirteenSingle
\caption{Rank distributions for Mexico City student preferences, single uniform priority distribution, 2013}%
\label{FIG:MEX_RANK_DIST_THIRTEEN_SING}%
\end{figure}
\begin{figure}[H]%
\PlotsMexRankDistFourteenSingle
\caption{Rank distributions for Mexico City student preferences, single uniform priority distribution, 2014}%
\label{FIG:MEX_RANK_DIST_FOURTEEEN_SING}%
\end{figure}
\begin{figure}[H]%
\PlotsMexRankDistTenMultiple
\caption{Rank distributions for Mexico City student preferences, multiple uniform priority distribution, 2010}%
\label{FIG:MEX_RANK_DIST_TEN_MULT}%
\end{figure}
\begin{figure}[H]%
\PlotsMexRankDistElevenMultiple
\caption{Rank distributions for Mexico City student preferences, multiple uniform priority distribution, 2011}%
\label{FIG:MEX_RANK_DIST_ELEVEN_MULT}%
\end{figure}
\begin{figure}[H]%
\PlotsMexRankDistTwelveMultiple
\caption{Rank distributions for Mexico City student preferences, multiple uniform priority distribution, 2012}%
\label{FIG:MEX_RANK_DIST_TWELVE_MULT}%
\end{figure}
\begin{figure}[H]%
\PlotsMexRankDistThirteenMultiple
\caption{Rank distributions for Mexico City student preferences, multiple uniform priority distribution, 2013}%
\label{FIG:MEX_RANK_DIST_THIRTEEN_MULT}%
\end{figure}
\begin{figure}[H]%
\PlotsMexRankDistFourteenMultiple
\caption{Rank distributions for Mexico City student preferences, multiple uniform priority distribution, 2014}%
\label{FIG:MEX_RANK_DIST_FOURTEEEN_MULT}%
\end{figure}

\section{Failure of Ordinal Efficiency for \BM, \ABM, and \DA}
\label{APP:ORDINAL_FAILURE}
The following example shows that neither $\BMU$, nor $\ABMU$, nor $\DAU$ are ordinally efficient. 
\begin{example}
\citet{Bogomolnaia2001ANewSolution} gave an example where $\DAU = \RSD$ is not ordinally efficient. 
Consider a setting with four students $N=\{1,\ldots,4\}$, fours schools $M=\{a,\ldots,d\}$ with a single seat each, and the preference profile $ P = (P_1,\ldots,P_4)$ with 
\begin{eqnarray*}
	P_1 & : & a \succ b \succ c \succ d, \\
	P_2 & : & a \succ c \succ b \succ d, \\
	P_3 & : & d \succ b \succ c \succ a, \\
	P_4 & : & d \succ c \succ b \succ a. \\
\end{eqnarray*}
On $ P$ both $\BMU$ and $\ABMU$ produce the same assignment. 
\begin{center}
\begin{tabular}{|c||c|c|c|c|}
	\hline
	Student & \hspace{0.55em}$a$\hspace{0.55em} & \hspace{0.55em}$b$\hspace{0.55em} & \hspace{0.55em}$c$\hspace{0.55em} & \hspace{0.55em}$d$\hspace{0.55em} \\
	\hline
	\hline
	1	& $1/2$ & $3/8$ & $1/8$ & $0$	\\
	\hline
	2 & $1/2$ & $1/8$ & $3/8$ & $0$	\\
	\hline
	3 & $0$   & $3/8$ & $1/8$ & $1/2$	 \\
	\hline
	4	& $0$   & $1/8$ & $3/8$ & $1/2$	\\
	\hline
\end{tabular}
\end{center}
This assignment is ordinally dominates by an assignment where students 1 and 2 exchange their probabilities for $b$ and $c$ and students 3 and 4 do the same. 
\begin{center}
\begin{tabular}{|c||c|c|c|c|}
	\hline
	Student & \hspace{0.55em}$a$\hspace{0.55em} & \hspace{0.55em}$b$\hspace{0.55em} & \hspace{0.55em}$c$\hspace{0.55em} & \hspace{0.55em}$d$\hspace{0.55em} \\
	\hline
	\hline
	1	& $1/2$ & $1/2$ & $0$ & $0$	\\
	\hline
	2 & $1/2$ & $0$ & $1/2$ & $0$	\\
	\hline
	3 & $0$   & $1/2$ & $0$ & $1/2$	 \\
	\hline
	4	& $0$   & $0$ & $1/2$ & $1/2$	\\
	\hline
\end{tabular}
\end{center}
\end{example}
\section{Failure of \BM, \ABM, and \DA\ to Lie on the Efficient Frontier}
\label{APP:EFF_FAILURE}
In this section we show that none of the mechanisms $\BMU$, $\ABMU$, and $\DAU$ lie on the efficient frontier among those mechanisms that share their axiomatic incentive properties. 
\begin{definition}[Efficient Frontier]  
\label{DEF:EFFICIENT_FRONTIER}
For some set $\Phi$ of mechanisms, we say that $\varphi \in \Phi$ is \emph{on the efficient frontier, subject to $\Phi$}, if $\varphi$ is not ordinally dominated by any other mechanism $\varphi'$ from $\Phi$ at all preference profiles where this dominance is strict for at least one preference profile.
\end{definition}
Before we formulate the results we need to define three more properties of mechanisms, namely symmetry, swap monotonicity, and upper invariance. 
\begin{definition}[Symmetry]  
\label{DEF:SYMMETRY}
A mechanism $\varphi^{\mathds{P}}$ is \emph{symmetric} if for any preferences profiles $ P \in \mathcal{P}^N$ with $P_i = P_{i'}$ for some $i \in N$ we have 
$\varphi_i^{\mathds{P}}( P) = \varphi_{i'}^{\mathds{P}}( P)$. 
\end{definition}
The \emph{neighborhood} $N_P$ of a preference order $P$ are the preference orders that differ from $P$ by a swap of two schools that are adjacent in the ranking under $P$. 

\begin{definition}[Swap Monotonicity]  
\label{DEF:SM}
A mechanism $\varphi^{\mathds{P}}$ is \emph{swap monotonic} if for 
any student $i\in N$, 
any preference profile $(P_i,P_{-i}) \in \mathcal{P}^N$, 
and any misreport $P_i' \in N_{P_i}$ from the neighborhood of $P_i$ with $P_i: a \succ b$ and $P_i': b \succ a$, 
we have that either $i$'s assignment does not change, or $\varphi^{\mathds{P}}_{i,a}(P_i,P_{-i}) < \varphi^{\mathds{P}}_{i,a}(P_i',P_{-i})$ and $\varphi^{\mathds{P}}_{i,b}(P_i,P_{-i}) > \varphi^{\mathds{P}}_{i,b}(P_i',P_{-i})$.
\end{definition}
\begin{definition}[Upper Invariance]  
\label{DEF:UI}
A mechanism $\varphi^{\mathds{P}}$ is \emph{upper invariant} if for 
any student $i\in N$, 
any preference profile $(P_i,P_{-i}) \in \mathcal{P}^N$, 
and any misreport $P_i' \in N_{P_i}$ from the neighborhood of $P_i$ with $P_i: a \succ b$ and $P_i': b \succ a$, 
we have that $i$'s assignment for schools from the upper contour set of $a$ does not change (i.e., $\varphi^{\mathds{P}}_{i,j}(P_i,P_{-i}) = \varphi^{\mathds{P}}_{i,j}(P_i',P_{-i})$ for all $j\in M$ with $P_i : j \succ a$).
\end{definition}

For $\DA^\UU = \text{RSD}$, \citet{Erdil2014SPStochasticAssignment} has already shown that this mechanism does not lie on the efficient frontier, subject to strategyproofness and symmetry when the total number of seats strictly exceeds the number of students. 
In other words, there exists a strategyproof, symmetric mechanism that ordinally dominates $\DAU$ at all preference profiles and this dominance is strict for at least one preference profile. 
Proposition \ref{PROP:EFF_BM_ABM} provides similar results for $\BMU$ and $\ABMU$.

\begin{proposition}
\begin{enumerate}
	\item $\BMU$ is not on the efficient frontier with respect to ordinal (or rank) dominance, subject to upper invariance and symmetry. 
	\item $\ABMU$ is not on the efficient frontier with respect to ordinal (or rank) dominance, subject to upper invariance, swap montonicity, and symmetry.
\end{enumerate}
\label{PROP:EFF_BM_ABM}
\end{proposition}

\begin{proof}
To see that $\BMU$ does not lie on the efficient frontier, we construct a mechanism that is upper invariant and ordinally dominates $\BM^\UU$. 
This mechanism, $\BM^+$, is essentially the same mechanism as $\BM^\UU$, except that the assignment is altered at certain preference profiles. 
Again, consider the setting with 4 students and 4 schools in unit capacity.
We say that a preference profile satisfies \emph{separable wants} if the schools and students can be renamed such that \begin{itemize}
	\setlength{\itemsep}{0pt}
	\item students 1 and 2 have first choice $a$, 
	\item students 3 and 4 have first choice $b$, 
	\item students 1 and 3 prefer $c$ to $d$, 
	\item and students 2 and 4 prefer $d$ to $c$. 
\end{itemize}
Formally,
\begin{eqnarray*}
	 P_1 & : &  a \succ \{b,c,d\} \text{ and } c \succ d, \\
	 P_2 & : &  a \succ \{b,c,d\} \text{ and } d \succ c, \\
	 P_3 & : &  b \succ \{a,c,d\} \text{ and } c \succ d, \\
	 P_4 & : &  b \succ \{a,c,d\} \text{ and } d \succ c.
\end{eqnarray*}
$\BM^+$ is the same as BM, except that the outcome is adjusted for preference profiles with separable wants.
Let
\begin{equation}
	\BM^+({ P}) = \left\{ \begin{array}{ll} \text{PS}({ P}) , & \text{if }{ P}\text{ satisfies separable wants}, \\
		\BM^\UU({ P}) , & \text{else},
	\end{array}\right.
\end{equation}
where PS denotes the Probabilistic Serial mechanism \citep{Bogomolnaia2001ANewSolution}.
At some preference profile ${ P}$ that satisfies separable wants, the assignment under PS (after appropriately renaming of the students and schools) is
\begin{equation}
\text{PS}({ P}) = \left(\begin{array}{cccc}
\frac{1}{2} & 0 & \frac{1}{2} & 0 \\
\frac{1}{2} & 0 & 0 & \frac{1}{2} \\
0 & \frac{1}{2} & \frac{1}{2} & 0 \\
0 & \frac{1}{2} & 0 & \frac{1}{2} 	
\end{array}\right).
\end{equation}
Observe that under $\BM^\UU$, $a$ is split equally between $1$ and $2$, and $b$ is split equally between $3$ and $4$.
Consequently, students $1$ and $2$ get no share of $b$ and students $3$ and $4$ get no share of $a$, just as under PS.
Among all assignments that distribute $a$ and $b$ in this way, student $1$ prefers the ones that give her higher probability at $c$. 
This is at most $\frac{1}{2}$, since she already receives $a$ with probability $\frac{1}{2}$. 
Similarly, students $2,3$, and $4$ prefer their respective assignment under PS to any other assignment that splits $a$ and $b$ in the same way as PS and $\BM^\UU$. 
Therefore, $\BM^+$ weakly ordinally dominates $\BM^\UU$, and the dominance is strict for preference profiles ${ P}$ with separable wants.

It remains to be shown that $\BM^+$ is upper invariant.
To verify this, we only need to consider the change in assignment that the mechanism prescribes if some student swaps two adjacent schools in its reported preference ordering.
Starting with any preference profile $ P$, the swap produces a new preference profile $ P'$. 
If neither $ P$ nor $ P'$ satisfy separable wants, the mechanism behave like $\BM^\UU$.

For swaps where at least one of the preference profiles satisfies separable wants, we can assume without loss of generality that this is $ P$. 
Such a swap will lead to a new preference profile $ P'$ and one of the following three cases:
\begin{enumerate}
	\setlength{\itemsep}{0pt}
	\item \label{case:still_separable_wants} The new preference profile satisfies separable wants.
	\item \label{case:fcp_changed} The composition of the first choices has changed.
	\item \label{case:same_fcp_not_separable_wants} The preference profile no longer satisfies separable wants, but the composition of the first choices has not changed.
\end{enumerate}
By symmetry, we can restrict our attention to student $1$, whose preference order satisfies
\begin{equation}
	 P_1: a \succ c \succ d \text{ and } a \succ b.
\end{equation}

Case (\ref{case:still_separable_wants}) implies that $ P'$ still satisfies separable wants with respect to the same mappings $\mu,\nu$. 
Thus, $\BM^+$ will not change the assignment, i.e., upper invariance is not violated.

In case (\ref{case:fcp_changed}), student 1 has a new first choice. 
If the new first choice is $b$, she will receive $b$ with probability $\frac{1}{3}$ and $a$ with probability $0$ under $\BM^\UU( P')$. 
If the new first choice is $c$ or $d$, the student will receive that that school with certainty under $\BM^\UU( P')$. 
Both changes are consistent with upper invariance.

Finally, in case (\ref{case:same_fcp_not_separable_wants}), the swap must involve $c$ and $d$, since this is the only way in which separable wants can be violated. 
Since $ P'$ violates separable wants, we have $\BM^+( P') = \BM^\UU( P')$, and therefore, $a$ will still be split equally between the students who rank it first, and the same is true for $b$. 
Thus, student $1$ will receive $\frac{1}{2}$ of $a$ and $0$ of $b$, which is the same as under $\BM^+( P)=\mathrm{PS}( P)$. 
The only change can affect the assignment for the schools $c$ and $d$. 
This is consistent with upper invariance.

\medskip
The proof that $\ABMU$ is not on the efficient frontier is analogous. 
We construct the mechanism $\ABM^+$ in the same way as $\BM^+$, i.e., we take $\ABM^\UU$ as a baseline mechanism but replace the outcomes for preference profiles with separable wants by the outcomes chosen by the PS mechanism. 

As for $\BM^+$, we consider a swap of two adjacent schools in the preference report of student 1, such that $ P$ satisfies separable wants. 

In case (\ref{case:still_separable_wants}), when the new profile also satisfies separable wants, the assignment does not change, which is consistent with upper invariance and swap monotonicity. 
 
In case (\ref{case:fcp_changed}), when the composition of first choices changes, student $1$ must have ranked her second choice first. In this case, she will receive this new first choice with probability $\frac{1}{3}$ and the prior first choice with probability $0$. 
This is also consistent with upper invariance and swap monotonicity. 

Finally, in case (\ref{case:same_fcp_not_separable_wants}), the swap must involve $c$ and $d$. 
She will still receive her first choice with probability $\frac{1}{2}$ and her second choice with probability $0$ (as in the proof for $\BM^\UU$). 
Therefore, her assignment for the school she brought down, can only decrease, and her assignment for the school she brought up can only increase, and both change by the same absolute value. 
This is consistent with upper invariance and swap monotonicity. 
\end{proof}

%
%
%
%
%
%
%
%
%
%
%
%
%
%
%
%
%
%
%
%
%
%
%
%
		%
%
%
%
%
%
%
%
%
%
%
%
%
%
%
%
%
%
%
%
%
%
%
%
%
%
%
%
%
%
%
%
%
%
%
%
%
%
%
%
%
%
%
%
%
%
%
%

\end{document}